\documentclass[lineno]{jfm}

\usepackage{graphicx}
\usepackage{epstopdf,epsfig}
\usepackage{newtxtext}
\usepackage{newtxmath}
\usepackage{natbib}
\usepackage{algorithm}
\usepackage{algpseudocode}
\usepackage{caption}
\usepackage{subcaption}
\usepackage{hyperref}
\hypersetup{
    colorlinks = true,
    urlcolor   = blue,
    citecolor  = black,
}

\newcommand{\RomanNumeralCaps}[1]

\title{Two-point Turbulence Closures in Physical Space}

\author{Noah Zambrano\aff{1}
  \corresp{\email{nzamb@umich.edu}}\and 
  Karthik Duraisamy\aff{1}
 }

\affiliation{\aff{1}Department of Aerospace Engineering, University of Michigan, Ann Arbor, MI 48103
}
\begin{document}
\maketitle
\begin{abstract}
This work presents a predictive two-point statistical closure framework for turbulence formulated in physical space. A closure model for ensemble-averaged, incompressible homogeneous isotropic turbulence (HIT) is developed as a starting point to demonstrate the viability of the approach in more general flows. The evolution equation for the longitudinal correlation function is derived in a discrete form, circumventing the need for a Fourier transformation. The formulation preserves the near-exact representation of the linear terms, a defining feature of rapid distortion theory. The closure of the nonlinear higher-order moments follows the phenomenological principles of the Eddy-Damped Quasi-Normal Markovian (EDQNM) model of \citet{orszag1970analytical}. Several key differences emerge from the physical-space treatment, including the need to evaluate a matrix exponential in the evolution equation and the appearance of triple integrals arising from the non-local nature of the pressure–Poisson equation. This framework naturally incorporates non-local length-scale information into the evolution of turbulence statistics. Verification of the physical-space two-point closure is performed by comparison with direct numerical simulations of statistically stationary forced HIT and with classical EDQNM predictions for decaying HIT. Finally, extensions to inhomogeneous and anisotropic turbulence are discussed, emphasizing advantages in applications where spectral methods are ill-conditioned, such as compressible flows with discontinuities.
\end{abstract}


\section{Introduction}
\label{sec:introduction}
In turbulence modeling, the Navier–Stokes equations serve as the foundation for deriving evolution equations of coarse-grained statistical quantities, which in turn provide closure for the coarse-grained equations themselves. The coarse-graining procedure, introduced to mitigate computational expense, depends on the choice of averaging or filtering operator—such as ensemble, temporal, or spatial averaging—selected according to the modeling framework. Application of this operator yields statistical moments that may be formulated as single- or multi-point, higher-order quantities. Owing to the prevalence of spatial derivatives in the governing equations, multi-point statistics offer a natural description of turbulent interactions. Nevertheless, most practical turbulence models employ single-point closures to reduce cost and analytical complexity, necessitating additional modeling assumptions to represent intrinsically multi-point phenomena.

Single-point closures often fail to accurately account for nonlocal interactions, anisotropy, and energy transfer across scales with sufficient fidelity. Two-point closures, such as the Eddy Damped Quasi‐Normal Markovian (EDQNM) \citep{orszag1970analytical}, Test‐Field Models (TFM) \citep{kraichnan1971almost}, Direct Interaction Approximation (DIA) \citep{kraichnan1959structure}, and related approaches, address many of these limitations, albeit demonstrated in canonical, and often homogeneous turbulence. These offer explicit representations of the two-point correlation functions which directly model scale‐to‐scale energy transfer. Two-point methods also allow representation of structural and wave effects – e.g. rotation and stratification – that are essentially inaccessible to simpler closures \citep{sagaut2008homogeneous}. Another advantage of two-point closures is the reduced reliance on empirical constants; in some formulations, closures can predict or constrain what would otherwise be \textit{ad hoc} parameters in single-point or lower-order models \citep{bos2006single}. Two-point closures also naturally reduce to rapid distortion theory in the rapid distortion/linear limit. This limiting behavior can be exploited in problems where the turbulence does not have time to interact with itself and the turbulence dynamics are significantly simplified, circumventing the closure problem in some special cases. Collectively, these advantages make two-point closures an appealing framework for flows where non‐locality, anisotropy, and details of energy transfer are important.

Historically, most two-point turbulence closures have been formulated in spectral space, where the governing equations for two-point velocity correlations simplify considerably due to statistical homogeneity. Classical models such as EDQNM and DIA operate on the energy spectrum and directly capture triadic interactions across wavenumbers \citep{orszag1970analytical,lesieur1978amortissement}. Spectral representations make the scale-to-scale transfer terms transparent and computationally efficient to evaluate. However, spectral methods tend to be limited to special canonical cases because of the mathematical limitations of the Fourier transform, and the vast majority of spectral models are formulated for incompressible flows. Very few works in the literature focus on variable-density cases, most notably the variable-density models by \citet{besnard1996spectral} and \citet{clark1995two}, and less on compressible flows, with the main example being the modeling work for compressible HIT by \citet{bertoglio2001two}. In a strict sense, application of Fourier transforms to problems with discontinuities, such as compressible flows with shocks, is ill-posed. \citet{canuto2007spectral} discusses the application of spectral methods to such problems, but almost all applications are spectral collocation methods which only transform the spatial derivatives to spectral space and not the dynamical variables. Without transformation of the variables, many of the spectral-based closure ideas originally developed for higher-order turbulence moments cannot be applied to this method. Therefore, this provides motivation for development of a two-point framework that enables translation of closure ideas to more complex discontinuous problems. A natural approach to this problem is to work exclusively in physical space.

Two-point closures are less common in physical space; yet they offer several advantages: they can be extended to inhomogeneous flows, more naturally preserve the locality of boundary conditions, and often provide more direct connections to experimentally measurable two-point statistics. Notable early developments include the work of \citet{besnard1996spectral}, who formulated a two-point closure in physical space for variable-density turbulence but ultimately utilized a Fourier transformation, and \citet{cambon2006anisotropic}, who advocated for real-space correlation closures as a complement to spectral models. More recently, scale-space transport equations based on the generalized K\'arm\'an–Howarth-Monin equations \citep{karman1938statistical,monin1959theory} have been derived explicitly in physical space, allowing one to track energy transfer simultaneously across spatial locations and separations. Most of these methods use the scale-space energy density function of \citet{hamba2015turbulent}. The work of \citet{arun2021scale} provides an example of how the scale-space transform can be used to compute two-point correlations in inhomogeneous compressible flows. However, these studies generally do not use the scale-space energy density function framework to derive closures of the moment equations; they are used as analysis/diagnostic tools for DNS data. To the authors' knowledge, there has not been development of a two-point physical space framework that can be systematically used to convert the relevant unclosed moment PDEs into ODEs and apply the closure ideas originally developed for spectral methods. The purpose of this work is to begin the development of a framework that can be used for prediction.

This work will consider the closure of incompressible HIT as a starting point to demonstrate the viability of the physical space two-point formulation and how it may be used in a predictive setting. HIT has been explored by many, including progress on two-point closures in physical space using the K\'arm\'an-Howarth \citep{karman1938statistical} equations. The early work of \citet{domaradzki1984simple} and \cite{oberlack1993closure} explored closure of the K\'arm\'an-Howarth equations with an eddy-viscosity model. Further work on modeling higher-order structure functions by \citet{thiesset2013karman} and \citet{djenidi2022karman}, to name a few, have also relied on eddy-viscosity type closures. All these methods rely on the K\'arm\'an-Howarth equations, which cannot be applied to general problems. To the authors' knowledge, there have not been attempts to adapt the ideas of EDQNM or DIA with the K\'arm\'an-Howarth equations. Thus, while closure of incompressible HIT is a common endeavor, the translation of the spectral closure ideas to a physical space two-point formulation is novel.

Instead of relying on a spectral transformation, a discrete approach is taken using finite differences to approximate spatial derivatives. For example, using a simple second-order central difference to discretize a spatial derivative in a third-order moment of some dynamical variable $u_i(\boldsymbol{x})$ evaluated at spatial points $\boldsymbol{x}$ and $\boldsymbol{y}$,
\begin{gather}\label{eq:general_derivative}
    \Big\langle u_i(\boldsymbol{x}) \frac{\partial u_j(\boldsymbol{x})}{\partial x_l}u_k(\boldsymbol{y})\Big\rangle \approx \Big\langle u_i(\boldsymbol{x}) \frac{\partial u_j(\boldsymbol{x}+\Delta x_l)}{2\Delta x_l}u_k(\boldsymbol{y})\Big\rangle -\Big\langle u_i(\boldsymbol{x})\frac{u_j(\boldsymbol{x}-\Delta x_l)}{2\Delta x_l}u_k(\boldsymbol{y})\Big\rangle,
\end{gather}
enables the spatial derivatives of higher-order moments to become a summation of neighboring moments. This highlights the most significant advantage of using two-point statistics over single-point statistics, as two-point information will inherently contain length-scale information, albeit truncated. This means that correlations with derivative terms do not require their own closure. Quantities such as
\begin{equation}\Big\langle \frac{\partial u_i}{\partial x_j}u_k\Big\rangle,\; \Big\langle u_i\frac{\partial u_j}{\partial x_l}u_k\Big\rangle
\end{equation}
will be closed if the higher-order two-point moments are known. Although this is not a significant advantage for simple problems such as incompressible HIT, it is important for more complex moment equations that arise in inhomogeneous anisotropic compressible turbulence, where many unclosed terms have spatial derivatives embedded into the higher-order moments. 

\section{Definitions and moment equations}
\label{sec:moment equations}
Definitions and moment equations are now discussed with generality in mind. Simplifications will be invoked later to make the problem more tractable but case-limited. A Reynolds-average coarse-graining operation is used, where average properties of the velocity field will be interpreted as ensemble averages over a large number of flow realizations, denoted by $\langle \;\rangle$. The velocity realizations are governed by the incompressible Navier-Stokes equations,
\begin{equation} \label{eq:momentum}
    \frac{\partial u_i(\boldsymbol{x})}{\partial t}=-u_j(\boldsymbol{x})\frac{\partial u_i(\boldsymbol{x})}{\partial x_j}-\frac{1}{\rho}\frac{\partial p(\boldsymbol{x})}{\partial x_i}+\nu \frac{\partial^2u_i(\boldsymbol{x})}{\partial x_j^2},
\end{equation}
\begin{equation}\label{eq:divergence-free}
    \frac{\partial u_i}{\partial x_i}=0,
\end{equation}
where $u_i(\boldsymbol{x})$ is the instantaneous $i^{\text{th}}$ velocity component evaluated at $\boldsymbol{x}$ in three-dimensional space, $p$ is the pressure, $\rho$ is the density, and $\nu$ is the kinematic viscosity. The pressure is closed by differentiating the momentum equation and solving the pressure-Poisson equation with the free-space Green function $1/4\pi |\boldsymbol{x}-\boldsymbol{x}'|$ \citep{durbin2011statistical},
\begin{equation}\label{eq:pressure}
    \frac{1}{\rho}\frac{\partial p(\boldsymbol{x})}{\partial x_i}=-\frac{1}{4\pi}\iiint_{-\infty}^{\infty}\frac{\partial}{\partial x'_i}\left(\frac{\partial u_j(\boldsymbol{x}')}{\partial x'_k}\frac{\partial u_k(\boldsymbol{x}')}{\partial x'_j}\frac{1}{|\boldsymbol{x}-\boldsymbol{x}'|}\right)d^3\boldsymbol{x}'.
\end{equation}
The general statistical quantities of interest are the multi-point moments, 
\begin{equation}\label{eq:phys_moments}
    R_{ij..k}(\boldsymbol{x},\boldsymbol{y},..,\boldsymbol{z})\triangleq\langle u_i(\boldsymbol{x})u_j(\boldsymbol{y})..u_k(\boldsymbol{z})\rangle.
\end{equation}
or more specifically the two-point velocity moment tensor, defined as
\begin{equation}\label{eq:two-point_vel_correlation}
    R_{ij}(\boldsymbol{x,y})\triangleq\langle u_i(\boldsymbol{x})u_j(\boldsymbol{y})\rangle.
\end{equation}
The second-order moment can also be written as a function of the distance vector between the two points by substituting $\boldsymbol{y}=\boldsymbol{x}+\boldsymbol{r}$. This work will only consider two-point statistical quantities but uses higher-order two-point moments in the closure scheme.

Now, the evolution equations for the two-point second- and third-order moments can be derived. To simplify notation, we introduce the multi-point spatial derivative of higher-order moments,
\begin{equation}\label{eq:general_derivative}
    \frac{\partial R_{ijk}(\boldsymbol{x,y,z})}{\partial y_l}\Big|_{y=x}\triangleq \Big\langle u_i(\boldsymbol{x}) \frac{\partial u_j(\boldsymbol{x})}{\partial x_l}u_k(\boldsymbol{z})\Big\rangle,
\end{equation}
where the conditional removes ambiguity by distinguishing which term in the moment the differential operator acts upon if two or more terms are evaluated at the same spatial point. The product rule is applied to $R_{ij}(\boldsymbol{x},\boldsymbol{y})$, and since the time derivative is a linear operator, we can use equation \ref{eq:momentum} to derive the second- and third-moment evolution equations,
\begin{multline} \label{eq:second-moment_evolution}
    \frac{\partial R_{ij}(\boldsymbol{x,y})}{\partial t}=-\frac{\partial R_{kij}(\boldsymbol{x,z,y})}{\partial z_k}\Big|_{z=x}-\frac{\partial R_{ijk}(\boldsymbol{x,z,y})}{\partial z_k}\Big|_{z=y}\\ -\frac{1}{\rho}\Bigl(\Big\langle \frac{\partial p(\boldsymbol{x})}{\partial x_i}u_j(\boldsymbol{y})\Big\rangle+\Big\langle \frac{\partial p(\boldsymbol{y})}{\partial y_j}u_i(\boldsymbol{x})\Big\rangle \Bigr)
    +\nu \frac{\partial^2 R_{ij}(\boldsymbol{x},\boldsymbol{y})}{\partial x_k^2}+\nu\frac{\partial^2 R_{ij}(\boldsymbol{x},\boldsymbol{y})}{\partial y_k^2}
\end{multline}
\begin{multline}  \label{eq:third-moment_evolution}
    \frac{\partial R_{ijk}(\boldsymbol{x,y,z})}{\partial t}=-\frac{\partial R_{lijk}(\boldsymbol{x,w,y,z})}{\partial w_l}\Big|_{w=x}-\frac{\partial R_{ljik}(\boldsymbol{y,w,x,z})}{\partial w_l}\Big|_{w=y}-\frac{\partial R_{lkij}(\boldsymbol{z,w,x,y})}{\partial w_l}\Big|_{w=z}\\
    -\frac{1}{\rho}\Bigl(\Big\langle \frac{\partial p(\boldsymbol{x})}{\partial x_i}u_j(\boldsymbol{y})u_k(\boldsymbol{z})\Big\rangle+\Big\langle \frac{\partial p(\boldsymbol{y})}{\partial y_j}u_i(\boldsymbol{x})u_k(\boldsymbol{z})\Big\rangle+\Big\langle \frac{\partial p(\boldsymbol{z})}{\partial z_k}u_i(\boldsymbol{x})u_j(\boldsymbol{y})\Big\rangle \Bigr)\\
    +\nu \frac{\partial^2 R_{ijk}(\boldsymbol{x},\boldsymbol{y},\boldsymbol{z})}{\partial x_k^2}+\nu\frac{\partial^2 R_{ijk}(\boldsymbol{x},\boldsymbol{y},\boldsymbol{z})}{\partial y_k^2}+\nu\frac{\partial^2 R_{ijk}(\boldsymbol{x},\boldsymbol{y},\boldsymbol{z})}{\partial z_k^2}
\end{multline}
From equations \ref{eq:second-moment_evolution}-\ref{eq:third-moment_evolution}, it is apparent that each moment equation requires information from the next higher-order moment. This is the closure problem, and is discussed in section \ref{sec:closure approximations}.
To simplify the problem, we assume that the turbulence is statistically homogeneous  and isotropic. Extension to general problems is discussed later in this paper. HIT properties allow for the two-point velocity moment to be written in terms of the one-dimensional longitudinal function $f(r)$ and the correlation distance vector $\boldsymbol{r}$ \citep{durbin2011statistical},
\begin{equation}\label{eq:longitudinal_function}
    R_{ij}(\boldsymbol{r})= u'^2\left(f(r)\delta_{ij}+\frac{r}{2}\frac{d f(r)}{\partial r}\left(\delta_{ij}-\frac{r_ir_j}{r^2}\right)\right).
\end{equation}
In HIT, the trace of the two-point velocity moment is the primary statistical quantity of interest. Hence, the dimensionality of the problem is reduced because $R_{ii}(r)$ is uniquely described by the one-dimensional longitudinal function and the correlation distance magnitude $r$. Imposing the constraints of incompressible HIT, the longitudinal function is related to the lateral function through
\begin{equation}\label{eq:lateral_function}
    g(r)= f(r)+\frac{r}{2}\frac{df(r)}{dr}.
\end{equation}
Finally, zero mean flow $\langle u_i\rangle=0$ and statistical homogeneity imply symmetry under permutation of points and indices and translational invariance,
\begin{equation}
     R_{ij..k}(\boldsymbol{x}+a,\boldsymbol{y}+a,..,\boldsymbol{z}+a)=R_{ij..k}(\boldsymbol{x},\boldsymbol{y},..,\boldsymbol{z}).
\end{equation}

\subsection{Spectral moments and transformation between spaces}
\label{sec:transformations}
Now, the spectral definitions are established strictly for homogeneous turbulence with zero mean flow $\langle u_i\rangle=0$. The multi-point $n^{\text{th}}$-order moments are related to the moments in Fourier space through the Fourier transform,
\begin{multline}
    \hat{R}_{ij..k}(\boldsymbol{\kappa},\boldsymbol{p}..\boldsymbol{q})\triangleq \langle \hat{u}_i(\boldsymbol{k})\hat{u}_j(\boldsymbol{p})...\hat{u}_k(\boldsymbol{q})\rangle= \\
    \frac{1}{(2\pi)^{3(n-1)}}\int  \langle u_i(\boldsymbol{x})u_j(\boldsymbol{y})...u_k(\boldsymbol{z})\rangle e^{-\mathrm{i}\boldsymbol{\kappa}\cdot \boldsymbol{x}-\mathrm{i}\boldsymbol{p}\cdot \boldsymbol{y}-...-\mathrm{i}\boldsymbol{q}\cdot \boldsymbol{z}}d^3\boldsymbol{x}d^3\boldsymbol{y}...d^3\boldsymbol{z}.
\end{multline}
Note that the trial function is the negative exponential instead of the positive one as seen in most Fourier transform definitions. We follow this notation to be consistent with \citet{orszag1970analytical}. In spectral space, the property $\boldsymbol{\kappa}+\boldsymbol{p}+...+\boldsymbol{q}=0$ arises from homogeneity due to translational invariance. This relation removes the functional dependence on the $n^{\textrm{th}}$ wavevector. This simplification is used throughout the remainder of the paper when working in spectral space. Another subtle restriction is that, in integrals over triadic interactions, we must avoid double-counting symmetric configurations. This implies that wavevector interactions must be non-degenerate, i.e. third-order spectral moments must obey the condition $\boldsymbol{\kappa}\neq-\boldsymbol{p}\neq\boldsymbol{q}\neq-\boldsymbol{\kappa}$ \citep{orszag1970analytical}.

HIT properties enable the representation of the spectral velocity second-moments using the wavenumber-dependent spherically symmetric energy spectrum $E(\kappa)$ and wavevector $\boldsymbol{\kappa}$,
\begin{equation}\label{eq:energy_spectrum}
    \langle \hat{u}_i(\boldsymbol{k})\hat{u}_j(\boldsymbol{p})\rangle=\frac{E(\kappa)}{4\pi\kappa^2}\left(\delta_{ij}-\frac{\kappa_i\kappa_j}{\kappa^2}\right)\delta(\boldsymbol{\kappa+p}).
\end{equation}
This is used to formally define the turbulent kinetic energy and turbulent dissipation rate,
\begin{gather}\label{eq:tke_dissipation}
    \frac{1}{2}q^2\triangleq\int_0^{\infty}E(\kappa)d\kappa=\frac{3}{2}u'^2,\\
    \epsilon\triangleq 2\nu\int_0^{\infty}\kappa^2E(\kappa)d\kappa=-15\nu \frac{d^2f_0}{dr^2}.
\end{gather}
and the turbulent dissipation rate. Now, following a similar procedure as in physical space, the moment equations are derived in spectral space,
\begin{equation} \label{eq:spec_second-moment_evolution}
    \left[\frac{d}{dt}+2\nu\kappa^2\right]\hat{R}_{ij}(\boldsymbol{\kappa},t)=\\
    -\frac{\mathrm{i}}{2}\int\left[P_{ikl}(\boldsymbol{\kappa})\hat{R}_{jkl}(-\boldsymbol{\kappa},\boldsymbol{p},t)+P_{jkl}(-\boldsymbol{\kappa})\hat{R}_{ikl}(\boldsymbol{\kappa},\boldsymbol{p},t)\right]d^3\boldsymbol{p}
\end{equation}
\begin{multline} \label{eq:spec_third-moment_evolution}
    \left[\frac{d}{dt}+2\nu(\kappa^2+p^2+q^2)\right]\hat{R}_{ijk}(\boldsymbol{\kappa},\boldsymbol{p},t)=-\frac{\mathrm{i}}{2}\int[P_{ilm}(\boldsymbol{\kappa})\hat{R}_{jklm}(\boldsymbol{p},\boldsymbol{q},\boldsymbol{r},t)\\
    +P_{jlm}(\boldsymbol{p})\hat{R}_{iklm}(\boldsymbol{\kappa},\boldsymbol{q},\boldsymbol{r},t)+P_{klm}(\boldsymbol{q})\hat{R}_{ijlm}(\boldsymbol{\kappa},\boldsymbol{p},\boldsymbol{r},t)]d^3\boldsymbol{r}\\
    -\mathrm{i}(P_{ilm}(\boldsymbol{\kappa})\hat{R}_{jl}(\boldsymbol{p},t)\hat{R}_{km}(\boldsymbol{q},t)
    +P_{jlm}(\boldsymbol{p})\hat{R}_{il}(\boldsymbol{\kappa},t)\hat{R}_{km}(\boldsymbol{q},t)\\
    +P_{klm}(\boldsymbol{q})\hat{R}_{il}(\boldsymbol{\kappa},t)\hat{R}_{jm}(\boldsymbol{p},t))=S_{ijk}(\boldsymbol{\kappa},\boldsymbol{p},t)
\end{multline}
where $\mathrm{i}=\sqrt{-1}$, and $P_{ijk}(\boldsymbol{\kappa})=\kappa_j(\delta_{ik}-\kappa_i\kappa_k/\kappa^2)+\kappa_k(\delta_{ij}-\kappa_i\kappa_j/\kappa^2)$ is the projection operator. Note the the terms involving products of $\hat{R}_{ij}(\boldsymbol{\kappa},t)$ arise due to the requirement that $\boldsymbol{\kappa}\neq-\boldsymbol{p}\neq\boldsymbol{q}\neq-\boldsymbol{\kappa}$ in the third-order spectral moment \citep{orszag1970analytical}.

A transformation between spectral and physical space must be established to compare turbulence statistics in each space. The energy spectrum is related to the longitudinal function through a Hankel-type transformation,
\begin{equation}\label{eq:E_to_R}
    u'^2f(r)=2\int_0^\infty E(\kappa)\frac{\sin(\kappa r)-\kappa r\cos(\kappa r)}{(\kappa r)^3}d\kappa.
\end{equation}
The energy spectrum from the two-point correlation is then given as
\begin{equation} \label{eq:R_to_E}
    E(\kappa)=\frac{1}{\pi}\int_0^\infty R_{ii}(r)\kappa r \sin(\kappa r)dr.
\end{equation}
These are oscillatory integrals involving Bessel-type kernels, so standard quadrature fails at high $\kappa$ or large $r$. To obtain an accurate Hankel transform over many decades of scale, the algorithm by \citet{talman1978numerical} is used.
\section{Closure approximations}
\label{sec:closure approximations}
Closure of the moment hierarchy is a deep and well-studied topic in turbulence. This work will focus on adapting the ideas originally developed by \citet{orszag1970analytical} in his EDQNM closure. This classical spectral closure was chosen because of its simplicity and phenomenological modeling that captures the Kolmogorov inertial range scaling. It is emphasized that the closure problem remains an active area of research, and many models and ideas may be used. Our goal is to demonstrate the viability of the two-point physical space framework described in section \ref{sec:physical space formulation} and provide an example of how it may be used in a predictive sense.
\subsection{Quasi-normality}
\label{sec:quasi-normality}
The first step towards closure of the moment equations is invoking the quasi-normal approximation \citep{Millionschtchikov1941}, which states that the fourth-order velocity moments can be approximated with a Gaussian distribution and can be neglected in the third-moment evolution equation. Gaussianity implies that fourth-order moments can be decomposed as a product of second-order moments,
\begin{multline}
    \langle u_i(\boldsymbol{x})u_j(\boldsymbol{y})u_k(\boldsymbol{z})u_l(\boldsymbol{w})\rangle\approx \\
    R_{ij}(\boldsymbol{x,y})R_{kl}(\boldsymbol{z,w})+R_{ik}(\boldsymbol{x,z})R_{jl}(\boldsymbol{y,w})+R_{il}(\boldsymbol{x,w})R_{jk}(\boldsymbol{y,z}).
\end{multline}
Another assumption is that the initial ensemble ($t=0$) is exactly Gaussian, which requires third-moments to be zero. This initial state receives some justification on the basis of the maximal randomness principle \citep{kraichnan1959structure}. Non-zero values of third-moments are developed in evolution so that the flow is not Gaussian after the initial instant. 
\subsection{Markovian modification}
\label{sec:markovian modification}
The Markovian modification enables analytical time integration of the spectral third-order moment equation by evaluating the spectra in equation \ref{eq:spec_third-moment_evolution} at current time $t$ instead of intermediate integration time $s$ (treating the spectra as quasi-constants in time). 
\begin{multline} \label{eq:spec_markovian}
    \left[\frac{d}{d t}+\nu(\kappa^2+p^2+q^2)\right]\hat{R}_{ijk}(\boldsymbol{\kappa},\boldsymbol{p},t)=S_{ijk}(\boldsymbol{\kappa},\boldsymbol{p},t)\\
    \longrightarrow \hat{R}_{ijk}(\boldsymbol{\kappa},\boldsymbol{p},t)\approx S_{ijk}(\boldsymbol{\kappa},\boldsymbol{p},t)\int_0^t e^{-\nu(\kappa^2+p^2+q^2)(t-s)}ds.
\end{multline}
where $S_{ijk}(\boldsymbol{\kappa},\boldsymbol{p},t)$ encompasses all nonlinear third-order moment terms. The Markovian modification is done \textit{a posteriori} and lacks fundamental justification \citep{orszag1973statistical}, but serves to significantly simplify the computations required to evolve the second-order moments.
\subsection{Eddy relaxation}
\label{sec:eddy relaxation}
Eddy relaxation is the phenomenological part of the EDQNM closure. By neglecting the effects of the fourth-order moments on the third-order moments, we have introduced unphysical behavior, lacking reversibility, realizability, and incorrect relaxation time in the energy spectrum. In EDQNM the relaxation time,
\begin{equation}\label{eq:spec_relaxation_time_undamped}
    \theta(\kappa,p,q,t)\triangleq\int_0^t e^{-\nu(\kappa^2+p^2+q^2)(t-s)}ds=\frac{1-e^{-\nu(\kappa^2+p^2+q^2)t}}{\nu(\kappa^2+p^2+q^2)},
\end{equation}
is multiplied by the nonlinear third-order moment $S_{ijk}(\boldsymbol{\kappa},\boldsymbol{p},t)$. \citet{orszag1970analytical} noted that the missing behavior of the fourth-order moments on the third-order moments can be accounted for by directly dampening $S_{ijk}(\boldsymbol{\kappa},\boldsymbol{p},t)$ with a modified ``eddy" relaxation time. The eddy relaxation time is created by modifying the molecular viscosity $\nu \kappa^2$ that appears in equation \ref{eq:spec_relaxation_time_undamped} to a scale-dependent eddy viscosity $\mu(\kappa)$. The typical formulation for the eddy viscosity \citep{andre1977influence} is
\begin{equation}\label{eq:spec_eddy_viscosity}
    \mu_k(t)\triangleq \nu \kappa^2+\lambda\sqrt{\int_0^\kappa p^2E(p,t)dp},\; \lambda =0.355.
\end{equation}
The final evolution equation for traditional EDQNM derived by \citet{orszag1970analytical} is
\begin{multline}\label{eq:EDQNM_equation}
    \left[\frac{d}{d t}+2\nu \kappa^2\right]E(\kappa,t)=
    \frac{1}{2}\iint_{\Delta}\theta(\kappa,p,q,t)\frac{\kappa}{pq}[2\kappa^2a(k,p,q)E(p,t)E(q,t)\\
    -p^2b(\kappa,p,q)E(\kappa,t)E(q,t)-q^2c(\kappa,p,q)E(\kappa,t)E(p,t)]dp dq,
\end{multline}
where the wavenumber triad geometrical coefficients are
\begin{equation}
    a(\kappa,p,q)=\frac{P_{ijk}(\boldsymbol{\kappa})P_{jl}(\boldsymbol{p})P_{km}(\boldsymbol{q})P_{ilm}(\boldsymbol{\kappa})}{4\kappa^2}=\frac{1}{2}(1-xyz-2y^2z^2),
\end{equation}
\begin{equation}
    b(\kappa,p,q)=-\frac{P_{ijk}(\boldsymbol{\kappa})P_{jl}(\boldsymbol{q})P_{kil}(\boldsymbol{p})}{2\kappa^2}=\frac{p}{\kappa}(xy+z^3),
\end{equation}
\begin{equation}
    c(\kappa,p,q)=b(\kappa,q,p)=2a(\kappa,p,q)-b(\kappa,p,q),
\end{equation}
and the cosine angles are
\begin{equation}
    z=\frac{\kappa^2+p^2-q^2}{2\kappa p},\;y=\frac{\kappa^2+q^2-p^2}{2\kappa q},\;x=\frac{p^2+q^2-\kappa^2}{2pq}.
\end{equation}

The relaxation time is evaluated and modified in the same way as \cite{andre1977influence} so that it is equal to $t$ at small times and equal to the true solution of $1/(\mu_k+\mu_p+\mu_q)$ at large times,
\begin{equation}
    \theta(\kappa,p,q,t)\approx\frac{t}{1+(\mu_k(t)+\mu_p(t)+\mu_q(t))t}.
\end{equation}

\section{Physical space formulation}
\label{sec:physical space formulation}
The key benefit to working exclusively in physical space is the avoidance of the Fourier transform to convert equation \ref{eq:momentum} from a PDE into an ODE. Instead of a Fourier transform, a discrete approach is taken using finite differences to approximate spatial derivatives. Derivatives can be evaluated with spectral accuracy using Fourier or Chebyshev collocation at the respective collocation nodes (assuming periodicity for Fourier or applying boundary conditions for Chebyshev \citep{canutospectral}). In general, this can be written in terms of matrix-vector multiplication,
\begin{equation}
    \mathbf{f}'=\boldsymbol{D}\mathbf{f},
\end{equation}
where lowercase bold symbols are $N\times 1$ column vectors and uppercase bold symbols are $N \times N$ matrices. $\mathbf{f}$ represents the state along a one-dimension grid with $N$ discrete points and $\boldsymbol{D}$ is the first derivative differentiation matrix, for example the Fourier collocation matrix \citep{canutospectral} is
\begin{equation}\label{eq:spectral_diff_mat}
D_{ij}=\left\{\begin{array}{ll}\frac{1}{2}(-1)^{i+j}\cot\left[\frac{(i-j)\pi}{N}\right], & i\neq j\\ 0 & i=j \end{array} \right.
\end{equation}
This spectrally accurate matrix is applicable only to periodic problems. Spectral collocation matrices are also dense matrices and thus fully nonlocal. We can approximate the derivatives semi-locally using standard differentiation matrices, for example
\begin{equation}\label{eq:standard_diff_mat}
   D_{ij} =
   \begin{cases}
   -\frac{1}{2h}, & j=i-1,\quad i=2,\dots,N-1,\\[1mm]
   \frac{1}{2h}, & j=i+1,\quad i=2,\dots,N-1,\\[1mm]
   0, & \text{otherwise (with modified rows near the boundaries)}.
   \end{cases}
\end{equation}
This kind of discretization will be applied to the evolution equations \ref{eq:second-moment_evolution}-\ref{eq:third-moment_evolution}. Finally, the linear and nonlinear terms of equation \ref{eq:second-moment_evolution} are explicitly defined as
\begin{equation} \label{eq:linear_terms}
    \frac{\partial R_{ij}(\boldsymbol{x,y})}{\partial t}\Big|_{\text{linear}}=
    \nu \frac{\partial^2 R_{ij}(\boldsymbol{x},\boldsymbol{y})}{\partial x_k^2}+\nu\frac{\partial^2 R_{ij}(\boldsymbol{x},\boldsymbol{y})}{\partial y_k^2}
\end{equation}
\begin{multline} \label{eq:nonlinear_terms}
    \frac{\partial R_{ij}(\boldsymbol{x,y})}{\partial t}\Big|_{\text{nonlinear}}=\underbrace{-\frac{\partial R_{kij}(\boldsymbol{x,z,y})}{\partial z_k}\Big|_{z=x}-\frac{\partial R_{ijk}(\boldsymbol{x,z,y})}{\partial z_k}\Big|_{z=y}}_{\text{advection terms}}\\ \underbrace{-\frac{1}{\rho}\Bigl(\Big\langle \frac{\partial p(\boldsymbol{x})}{\partial x_i}u_j(\boldsymbol{y})\Big\rangle+\Big\langle \frac{\partial p(\boldsymbol{y})}{\partial y_j}u_i(\boldsymbol{x})\Big\rangle \Bigr)}_{\text{pressure terms}}.
\end{multline}
We will investigate the application of the spatial discretization to both linear and nonlinear terms.
\subsection{Linear terms}
\label{sec:linear terms}
 We start with the linear term (viscous diffusion). The tensorial ODE is simplified by contracting indices, transforming to spherical coordinates using a change of variables $r=|\boldsymbol{y}-\boldsymbol{x}|$, and applying product rule:
\begin{equation}\label{eq:linear_radial}
    \frac{\partial R_{ii}(r)}{\partial t}\Big|_{\text{linear}}=2\nu\left( \frac{\partial^2 R_{ii}(r)}{\partial r^2}+\frac{2}{r}\frac{\partial R_{ii}(r)}{\partial r}\right).
\end{equation}
This is fully closed on the collocation grid and does not require modeling. Equation \ref{eq:linear_radial} is rewritten in terms of the longitudinal function using equation \ref{eq:longitudinal_function}. The evolution equation becomes
\begin{equation}\label{eq:linear_longitudinal_evolution}
    \left(3+r\frac{\partial}{\partial r}\right)\frac{\partial (u'^2f(r,t))}{\partial t}\Big|_{\text{linear}}=2\nu\underbrace{\left(\frac{\partial^2 }{\partial r^2}+\frac{2}{r}\frac{\partial}{\partial r}\right)}_{\text{radial Laplacian}}\left(3+r\frac{\partial}{\partial r}\right)u'^2f(r,t).
\end{equation}
The double-derivative operator is the radial Laplacian. Conversion of this PDE into an ODE is accomplished by discretizing the Laplacian operator. The discrete radial Laplacian is represented concisely by the radial Laplacian matrix $\boldsymbol{L}$. Finally, equation \ref{eq:linear_longitudinal_evolution} is rewritten in terms of $f(r)$ by inverting the differentiation matrix and obtaining the standard matrix-vector differential equation,
\begin{equation}
    \frac{d \boldsymbol{f}}{d t}\Big|_{\text{linear}} = \Bigl[3\boldsymbol{I} + \boldsymbol{R}\boldsymbol{D}\Bigr]^{-1}2\nu\boldsymbol{L}\Bigl[3\boldsymbol{I} + \boldsymbol{R}\boldsymbol{D}\Bigr]\boldsymbol{f},
\end{equation}
where $\boldsymbol{f}$ is $u'^2f(r)$ evaluated at each discrete $r$ value, $\boldsymbol{R}$ is a diagonal matrix with $r$ values along the diagonal, $\boldsymbol{I}$ is the identity matrix, and $\boldsymbol{D}$ is the first-derivative differentiation matrix. The general solution to the first-order matrix ODE is the matrix-exponential equation
\begin{equation}\label{eq:linear_soln}
    \boldsymbol{f}(t)=\exp{\Bigl(\Bigl[3\boldsymbol{I} + \boldsymbol{R}\boldsymbol{D}\Bigr]^{-1}\,2\nu\boldsymbol{L}\Bigl[3\boldsymbol{I} + \boldsymbol{R}\boldsymbol{D}\Bigr]t\Bigr)}\boldsymbol{f}(0).
\end{equation}
This is related to the lateral function by 
\begin{equation}\label{eq:mat-vec_lateral}
    \boldsymbol{g}=\boldsymbol{f}+\frac{1}{2}\boldsymbol{RDf}.
\end{equation} 
Note that any expression with $r$ in the denominator (such as $\boldsymbol{L}$) becomes undefined if $r=0$. To avoid this, the limit as $r\rightarrow0$ must be taken,
\begin{equation}\label{eq:limit}
    \lim_{r\rightarrow 0}\frac{f'(r)}{r}=f''(0).
\end{equation}
Another subtlety is that the numerical inversion of the matrix does not necessarily preserve the boundary conditions in the differentiation matrix, boundary conditions must be enforced post-inversion to avoid unphysical accumulation of numerical error.
\subsection{Nonlinear terms}
\label{sec:nonlinear terms}
The nonlinear terms are also discretized and converted into matrix-vector form, with the key difference that the higher-order moments must be solved through a third-order moment evolution equation. With the assumptions from EDQNM, this is done analytically, such that there is no need to evolve the third-order moment equation \ref{eq:third-moment_evolution}. However, it is emphasized that this is not generally possible. For most problems, the evolution equation of interest shall be the $(n-1)^{\textrm{th}}$-order moments, where the $n^{\textrm{th}}$-order moments are closed with closure approximations.

We begin with the advection terms, noting that upon contraction of indices, the two advection terms in equation \ref{eq:nonlinear_terms} are equivalent. This is because the expression should be invariant of direction in HIT and the derivatives of each velocity component are the same, so $\frac{\partial R_{kii}(\boldsymbol{x,z,y})}{\partial z_k}|_{z=x}=\frac{\partial R_{iik}(\boldsymbol{x,z,y})}{\partial z_k}|_{z=y}$. With this, we derive the corresponding third-order moment evolution equation,
\begin{multline}\label{eq:triple_moment_adv}
    \frac{\partial}{\partial t} \frac{\partial R_{kii}(\boldsymbol{x,z,y})}{\partial z_k}\Big|_{z=x}=M_{a,a}(\boldsymbol{x,y)}+M_{a,p}(\boldsymbol{x,y})\\
+\nu\frac{\partial R_{kii}(\boldsymbol{x,z,y})}{\partial z_k}\Big|_{z=x}\left(\frac{\partial^2}{\partial x_l^2}+\frac{\partial^2 }{\partial y_l^2}\right),
\end{multline}
with the nonlinear fourth-order moment terms, 
\begin{multline}\label{eq:markovian_a_a}
    M_{a,a}(\boldsymbol{x,y})\triangleq -\frac{\partial ^2R_{jkii}(\boldsymbol{x,z,w,y})}{\partial z_j\partial w_k}\Big|_{z,w=x}-\frac{\partial ^2R_{jiki}(\boldsymbol{x,z,x,y})}{\partial z_j\partial z_k}\Big|_{z=x}\\-\frac{\partial ^2R_{kiji}(\boldsymbol{x,z,w,y})}{\partial z_j\partial w_k}\Big|_{z,w=x}-\frac{\partial ^2R_{kiij}(\boldsymbol{x,w,z,y})}{\partial z_j\partial w_k}\Big|_{w=x,z=y},
\end{multline}
\begin{multline}\label{eq:markovian_a_p}
     M_{a,p}(\boldsymbol{x,y})\triangleq -\frac{1}{\rho}\Bigg[ \Big\langle\frac{\partial p(\boldsymbol{x})}{\partial x_k} \frac{\partial u_i(\boldsymbol{x})}{\partial x_k}u_i(\boldsymbol{y})\Big\rangle\\
     + \Big\langle\frac{\partial ^2 p(\boldsymbol{x})}{\partial x_i\partial x_k} u_k(\boldsymbol{x})u_i(\boldsymbol{y})\Big\rangle + \Big\langle\frac{\partial p(\boldsymbol{y})}{\partial y_i} \frac{\partial u_i(\boldsymbol{x})}{\partial x_k}u_k(\boldsymbol{x})\Big\rangle \Bigg],
\end{multline}
where subscript $a$ refers to the advection term and $p$ the pressure terms. The first subscript position corresponds to the second-moment evolution contribution and the second position as the triple-moment contribution.

The same can be done for the pressure terms in equation \ref{eq:nonlinear_terms}. However, it was found that by enforcing the divergence-free condition in the third-order moment equation for the advection term (equation \ref{eq:triple_moment_adv}), solving for the pressure terms was not required. The reasoning behind this is given in appendix \ref{sec:negligible pressure}. Now, we transform equation \ref{eq:triple_moment_adv} to spherical coordinates, discretize the derivatives, and rewrite in matrix-vector form. This can be written as a general two-point third-order moment equation in matrix-vector form,
\begin{equation}\label{eq:mat_vec_third_moment_evolution}
    \frac{d \boldsymbol{s}}{d t} 
= \boldsymbol{m}+2\nu \boldsymbol{L}\boldsymbol{s},
\end{equation}
where $\boldsymbol{s}$ is any general third-order moment evaluated at two points at various distances $r=|\boldsymbol{y}-\boldsymbol{x}|$, for example from equation \ref{eq:triple_moment_adv}, $\boldsymbol{s}=(\partial/\partial z_k) R_{kii}(\boldsymbol{x,z,y})|_{z=x}$. $\boldsymbol{m}$ is the respective nonlinear fourth-order moment terms. We can finally write equation \ref{eq:second-moment_evolution} in matrix-vector form after index contraction, coordinate transformation, and $f(r)$ substitution:
\begin{equation}\label{eq:mat_vec_second_moment_evolution}
    \frac{\partial \boldsymbol{f}}{\partial t} = \Bigl[3\boldsymbol{I} + \boldsymbol{R}\boldsymbol{D}\Bigr]^{-1}\left(-2\boldsymbol{s}^{(a)}-2\boldsymbol{s}^{(p)}+2\nu\boldsymbol{L}\Bigl[3\boldsymbol{I} + \boldsymbol{R}\boldsymbol{D}\Bigr]\boldsymbol{f}\right),
\end{equation}
where $\boldsymbol{s}^{(a)}$ is the advection term and $\boldsymbol{s}^{(p)}$ represent pressure terms from equation \ref{eq:nonlinear_terms}. If the pressure terms can be ignored $\boldsymbol{s}^{(p)}= \boldsymbol{0}$

While equation \ref{eq:mat_vec_third_moment_evolution} can be solved analytically using the assumptions from EDQNM, it is valuable to investigate how one would solve such an equation without the Markovian assumption on the fourth-order moments. It is also valuable to see how the nonlinear term impacts the solution of $\boldsymbol{s}$. Rather than forming a matrix exponential, similar to what was done in the linear terms, we note that equation \ref{eq:mat_vec_third_moment_evolution} is a forced diffusion equation and can be solved using the 3D heat kernel Green's function. The heat kernel is simplified to radial form under isotropy,
\begin{equation}\label{eq:radial_heat_kernel}
    G(r,r';\tau)=\frac{r'}{r\sqrt{4\nu\pi\tau}}\exp\left(-\frac{r^2+r'^2}{4\nu\tau}\right)\sinh\left(\frac{rr'}{2\nu\tau}\right).
\end{equation}
With zero initial condition for $\boldsymbol{s}$, Duhamel's principle is applied to give the closed-form solution
\begin{equation}\label{eq:s_soln_greens}
    s(r,t)=\int^t_0d\tau\int^{\infty}_0dr'G(r,r';\tau)M(r',t-\tau)r'^2.
\end{equation}
This is a radial Gaussian-convolution in time.
\subsection{Markovian modification and quasi-normality in physical space}
\label{sec:markovian in physical space}
We will adapt the ideas of EDQNM and apply Markovian modification to the nonlinear fourth-order moments. This enables analytical time-integration of equation \ref{eq:mat_vec_third_moment_evolution} and produces the general solution,
\begin{equation}\label{eq:mat_vec_third_moment}
    \boldsymbol{s}\approx \int^t_0e^{2\nu\boldsymbol{L}(t-\tau)}d\tau\boldsymbol{m}=(2\nu\boldsymbol{L})^{-1}(e^{2\nu\boldsymbol{L}t}-\boldsymbol{I})\boldsymbol{m},
\end{equation}
where the $e^{2\nu\boldsymbol{L}t}\boldsymbol{s}_0$ term is omitted from the solution as $\boldsymbol{s}_0=0$ for the cases presented in this work. 

The Markovian terms are now transformed and expressed in terms of the longitudinal function $f(r)$ by applying quasi-normality. For brevity, we only give the final solutions for $M_{a,a}(\boldsymbol{x},\boldsymbol{y})$ and $M_{a,p}(\boldsymbol{x},\boldsymbol{y})$. The full derivations are given in appendix \ref{sec:markovian term derivations}. $M_{a,a}(\boldsymbol{x},\boldsymbol{y})$ analytically reduces to
\begin{equation}\label{eq:m_a_a}
M_{a,a}(\boldsymbol{x},\boldsymbol{y})=u'^4\left[\left(rf'''(r)+7f''(r)+\frac{8}{r}f'(r)\right)(f(r)-1)+2.5f'(r)^2\right].
\end{equation}
For $M_{a,p}(\boldsymbol{x},\boldsymbol{y})$, the pressure-Poisson equation \ref{eq:pressure} is used to express the pressure in terms of velocity. Use of the pressure-Poisson equation introduces non-locality through the spherical integral, which must be solved numerically. For example, the first term of $M_{a,p}(\boldsymbol{x},\boldsymbol{y})$ is rewritten as
\begin{equation}\label{eq:greens_sol_map}
    -\frac{1}{\rho} \Bigg\langle\frac{\partial p(\boldsymbol{x})}{\partial x_k} \frac{\partial u_i(\boldsymbol{x})}{\partial x_k}u_i(\boldsymbol{y})\Bigg\rangle=-\int \left\langle \frac{\partial u_l(\boldsymbol{z})}{\partial z_j} \frac{\partial u_j(\boldsymbol{z})}{\partial z_l} \frac{\partial u_i(\boldsymbol{x})}{\partial x_k} u_i(\boldsymbol{y}) \right\rangle \frac{\partial G(\boldsymbol{x} - \boldsymbol{z})}{\partial x_k} \, d^3\boldsymbol{z}.
\end{equation}
The fourth-order moments will depend on distance and direction. We define three different distances based on four spatial positions, $\boldsymbol{r}=\boldsymbol{y-x},\boldsymbol{r}'=\boldsymbol{z-x}$, and $\boldsymbol{w-z}=\boldsymbol{r}''$. The general form of the integral for some function $\mathcal{F}(\boldsymbol{x,y,z})$ is, assuming $\boldsymbol{x}=0$ and transforming into spherical coordinates,
\begin{equation}\label{eq:spherical_integral}
    \int \mathcal{F}(\boldsymbol{x,y,z})d^3\boldsymbol{z}=\int^{\phi=2\pi}_{\phi=0}\int^{\theta=\pi}_{\theta=0}\int^{r=\infty}_{r=0}\mathcal{F}(r,r',|r-r'|,\theta,\phi)r'^2\sin\theta dr'd\theta d\phi.
\end{equation}
The steps for rewriting the fourth-order moments in terms of products of $f(r)$ are given in detail in appendix \ref{sec:markovian term derivations}. The final expression is in terms of $f(r),f(r'),$ and $f(|\boldsymbol{r}-\boldsymbol{r}'|)$. A critical observation is that $f(|\boldsymbol{r}-\boldsymbol{r}'|)$ depends on the orientation of $\boldsymbol{r'}$ and therefore the orientation cannot be analytically integrated. The final solution for $M_{a,p}(\boldsymbol{x},\boldsymbol{y})$ is lengthy and can be found in the appendix \ref{sec:markovian term derivations}. The pressure term is the most complex and expensive part of the formulation and is unique to incompressible flows.
\subsection{Eddy relaxation in physical space}
\label{sec:eddy relaxation in physical space}
The purpose of eddy damping is to relax the two-point third-order moment terms in the second-order moment evolution equation. This accounts for the missing effects of the fourth-order moments, where coherence is destroyed by non-linear scrambling \citep{orszag1973statistical}. \citet{orszag1970analytical} implemented eddy-damping by changing the molecular viscosity to an eddy viscosity in the memory integral of the third-order moment solutions. In physical space, the analogue of this is achieved by changing $\nu\boldsymbol{L}$ to a phenomenological eddy-viscosity $\boldsymbol{\mu}$ in the matrix exponential. This modification is readily seen by recalling the integral solution for two-point third-order moments,
\begin{equation}\label{eq:eddy_damping_location}
\boldsymbol{s}\approx \int^t_0e^{2\boldsymbol{\mu}(t-\tau)}d\tau\boldsymbol{m}=(2\boldsymbol{\mu})^{-1}(e^{2\boldsymbol{\mu}t}-\boldsymbol{I})\boldsymbol{m}.
\end{equation}
The general expression for the eddy-viscosity matrix is
\begin{equation} \label{eq:eddy_damping_form}
    \boldsymbol{\mu}\triangleq\nu\boldsymbol{L}+\boldsymbol{\nu}_t,\; [\nu\boldsymbol{L},\boldsymbol{\nu}_t]=0,
\end{equation}
where the two matrices are restricted to be commutable such that the impact of the damping is directly proportional to the exponential of the Laplacian. We now develop an analogous formulation by comparing directly to the general solution of two-point third-order moments in traditional EDQNM, given by equation \ref{eq:spec_markovian}. In the most basic form, as given by \citet{orszag1970analytical}, EDQNM uses the eddy-viscosity $\mu(\kappa)=C_\mu\epsilon^{1/3}\kappa^{2/3}$. The Fourier-space equivalent of the Laplacian matrix operator is $-\kappa^2$, and thus, by comparison and unit analysis, the eddy viscosity in physical space will take the form
\begin{equation}\label{eq:eddy_damping_first_form}
\boldsymbol{\nu}_t\triangleq C_{\mu}\epsilon^{1/3}(-\boldsymbol{L})^{1/3}.
\end{equation}
At small correlation lengths $\Delta r \gtrsim \lambda$, the fractional Laplacian operator tends to under-damp due to numerical errors originating from the small discretization. A simpler formulation is created by computing the damping locally through a diagonal matrix, inspired by the most widely used eddy damping form given by equation \ref{eq:spec_eddy_viscosity} \citep{andre1977influence,lesieur2008introduction}
\begin{equation}\label{eq:eddy_damping_final_form}
    \boldsymbol{\nu}_t\triangleq C_{\mu}\sqrt{S_2(r)}/r.
\end{equation}
This formulation is more robust near the limit $r\rightarrow 0$ because it explicitly constructs the correct damping rate $\delta u'(r)/r \sim u'/\lambda $. The second-order structure function $S_2(r)=2u'^2[1-f(r)]$ is used to provide information on the scale-dependent velocity increment, the ``localized" part of the eddy viscosity model form. This model form is inspired by the closure devised by \citet{oberlack1993closure}. Note that at $r=0$, a Taylor-series expansion is used to evaluate $\boldsymbol{\nu}_t$. A tuneable coefficient of $C_{\mu}=-3.5$ is used.
\subsection{Analogy to K\'arm\'an-Howarth equation}
\label{sec:analogy to karman-howarth equation}
The matrix-vector equation derived for the longitudinal function, equation \ref{eq:mat_vec_second_moment_evolution}, is analogous to the classical K\'arm\'an-Howarth equation \citep{karman1938statistical}
\begin{equation}\label{eq:karman-howarth}
    \frac{\partial u'^2f}{\partial t}=\frac{u'^3}{r^4}\frac{\partial }{\partial r}\left(r^4h\right)+\frac{2\nu u'^2}{r^4}\frac{\partial }{\partial r}\left(r^4\frac{\partial f}{\partial r}\right),
\end{equation}
where $h$ is the unclosed third-order moment. The viscous term is identical to the Laplacian matrix-vector term in equation \ref{eq:mat_vec_second_moment_evolution}, while $\Bigl[3\boldsymbol{I} + \boldsymbol{R}\boldsymbol{D}\Bigr]^{-1}(-\boldsymbol{s}^{(a)}-\boldsymbol{s}^{(p)})$ corresponds to the term involving $h$ in the K\'arm\'an-Howarth equation. Several notable works \citep{domaradzki1984simple,oberlack1993closure} have closed the K\'arm\'an-Howarth equation using an eddy-viscosity type of closure, which is written as
\begin{equation}
    \frac{\partial u'^2f}{\partial t}=\frac{2 u'^2}{r^4}\frac{\partial }{\partial r}\left(\nu_Tr^4\frac{\partial f}{\partial r}\right),
\end{equation}
where $\nu_T=\nu+A(r,t)$ is the scale-dependent eddy-viscosity. The closure model is established through $A(r,t)$, and has the simple form (derived from Kolmogorov's similarity theory) $A(r,t)=C_{\mu}\epsilon^{1/3}r^{4/3}$. \citet{domaradzki1984simple} originally proposed this type of eddy-viscosity closure. It is simple and reproduces the Kolmogorov scaling in the inertial subrange, however the approach is less explicit about nonlocal triadic effects. \citet{oberlack1993closure} also use the eddy-viscosity formulation for $h$ but enforces limiting behavior at large- and small-$r$ values through constraints of continuity and incompressibility.

More recent work has developed closures for the third-order structure function \citep{thiesset2013karman,djenidi2021modeling,djenidi2022karman}. The K\'arm\'an-Howarth equation is rewritten in terms of the second- and third-order structure functions $S_2,S_3$
\begin{equation}\label{eq:karma-howarth_strucutre_functions}
    6\nu\frac{\partial S_2}{\partial r}-S_3-\frac{3}{r^4}\int_0^rr'^4\frac{\partial S_2}{\partial t}dr'=\frac{4}{5}\overline{\epsilon} r,
\end{equation}
where $S_3$ is again modeled with an eddy-viscosity, $S_3=-\nu_T\partial S_2/\partial r$. Existing models for the third-order structure function and the general third-order moment $h$ are formulated specifically for use with the K\'arm\'an–Howarth equation, and no straightforward extension to anisotropic or inhomogeneous turbulence currently exists. The closure model developed in this work is likewise restricted to homogeneous isotropic turbulence (HIT); however, the underlying concept---closing the hierarchy at the level of the fourth-order moments while explicitly evolving the third-order moments---can, in principle, be generalized. The points at which the assumptions of homogeneity and isotropy enter the formulation are clearly identifiable through the substitution of the longitudinal correlation function for $R_{ii}(\boldsymbol{r})$. Further discussion on extending the present framework to more general flow configurations is provided in Section \ref{sec:extension to general problems}.
\section{Verification}
\label{sec:verification}
In this section, numerical experiments are performed to demonstrate that our physical space model is able to replicate the predictive abilities of spectral EDQNM. First, the linearized physical space closure is compared to the linearized part of EDQNM for a decaying initial spectrum. This verifies the equivalence of the matrix exponential Laplacian with spectral differentiation. Next, the full nonlinear formulation is checked with decaying HIT. This verifies a correct implementation of the nonlinear parts of the formulation. Finally, the new formulation is verified against forced HIT. This is used to validate the physical space model against DNS data and compare higher-order moments.
\subsection{Numerical considerations}
\label{sec:numerical}
Implementation of the physical space formulation relies on accurate numerical computation of the matrix exponential, correct and stable time integration methods, and adequate grid spacing. In addition, several non-dimensional parameters are used to characterize the HIT cases. The intensity of the turbulence is characterized by the Taylor Reynolds number and integral-scale Reynolds number,
\begin{equation}\label{eq:taylor_reynolds_number}
    Re_{\lambda}\triangleq\frac{u'\lambda}{\nu},\; Re_{\ell}\triangleq\frac{u'}{\nu \kappa_I},
\end{equation}
where $\kappa_I$ is the wavenumber where the energy spectrum peaks, set to unity unless otherwise stated. The two length-scales of interest are the Taylor microscale and the integral length-scale,
\begin{gather}\label{eq:length_scales}
    \lambda\triangleq\sqrt{\frac{15\nu u'^2}{\epsilon}},\\\ell\triangleq\frac{\pi}{2u'^2}\int_0^{\infty}\frac{E(\kappa)}{\kappa}d\kappa=\int_0^{\infty}f(r)dr.
\end{gather}
In spectral space, the maximum wavenumber determines the integral-scale Reynolds number $Re_{\ell}$. The maximum wavenumber suggested by \citet{lesieur2008introduction}, \citet{andre1977influence} should be found through the relation $\kappa_{\text{max}}=8\kappa_IRe_{\ell}^{3/4}$. The wavenumber grid is geometrically spaced,
\begin{equation}\label{eq:grid_discretization}
    \kappa_i=\kappa_0(2^{i/F}),\,i=0,1,...,N-1,
\end{equation}
with $\kappa_0=1/4$ and $F=4$ selected for all cases presented in this work. For example, 65 wavenumber points with this spacing gives $Re_{\ell}=26000$, 50 points gives $Re_{\ell}=813$, etc. Numerical integration of the convolution integrals in EDQNM is carried out using the method described by \citet{leith1971atmospheric}. 

The correlation distance $r$ was specified \textit{a priori}. The $r$-grid was constructed to capture spatial scales consistent with those represented in the traditional EDQNM model. The grid was initialized at $r=0$ to enable evaluation of single-point statistics. The maximum correlation distance was chosen to exceed the spatial scale associated with the smallest resolved wavenumber, thereby mitigating oscillations arising during the transformation between physical and spectral spaces. A geometrically spaced grid, refined toward smaller $r$-values, was employed to accurately resolve derivatives of the longitudinal correlation function up to third order. Unless otherwise stated, 100 grid points were used. This refinement at small $r$ was found to be essential when the energy spectrum exhibited a well-developed inertial range. The grid points are distributed according to 
\begin{gather}\label{eq:grid_discretization}
    r_i=r_{min}\left[(r_{max}/r_{min})^{(1/N-2)}\right]^{i-1},\;i=1,2,...,N-1,\\
    r_0=0,\;r_{min}=\pi/\kappa_{max},\;r_{max}=\pi F/(\kappa_{min}\log{2}).
\end{gather}
The nonlocal spherical integral for the pressure term was numerically approximated using a 50-point Gaussian quadrature for angle integration and 100 geometrically-spaced points for distances $r,r'$. Neumann boundary conditions are applied to both ends of the grid. Note that the exact boundary condition at the largest value of $r$ is not generally known, but if $r_{max}$ is large enough, then a Neumann boundary condition is a reasonable approximation. Differentiation matrices are made using radial basis functions. This is detailed in the appendix \ref{sec:rbf}. The physical space evolution equations require matrix exponential evaluation. For this, we follow the methodology outlined by \citet{higham2005scaling} and \citet{al2010new}. To initialize the spectrum, one must specify $\kappa_0$, $\kappa_I$, $F$, and the number of points for both $r$ and $\kappa$ discretization.

When the equations cannot be solved analytically in time, a forward Euler time-stepping scheme was used for spectral methods while a mixed implicit-explicit scheme was used for the physical space method. Traditional explicit schemes require that the time step $\Delta t$ is small enough to satisfy the stability criteria \citep{lesieur2008introduction},
\begin{equation}\label{eq:stability_spec}
    \Delta t\leq \frac{1}{\nu \max(\kappa)^2}.
\end{equation}
The stiff part of the ODE is the viscous term. Similar eigenvalue stability analysis can be applied to the physical space formulation and yields the criteria
\begin{equation}\label{eq:stability_phys}
    \Delta t\leq \frac{1}{\max_j|{\text{Re}(\lambda_j(\boldsymbol{A}))}|}.
\end{equation}
where $\lambda_j(\boldsymbol{A})$ is the maximum eigenvalue of the matrix appearing in the viscous term, $\boldsymbol{A}=2\nu\Bigl[3\boldsymbol{I} + \boldsymbol{R}\boldsymbol{D}\Bigr]^{-1}\boldsymbol{L}\Bigl[3\boldsymbol{I} + \boldsymbol{R}\boldsymbol{D}\Bigr]$. For the geometric spacing, the $r$-points near $r=0$ are on the order of $1\times10^{-4}$, creating a very stiff Laplacian matrix which in turn gives very large negative eigenvalues and prohibitively small timestep size. To speed up the simulations, an implicit-explicit (IMEX) scheme is used, where the nonlinear terms are evaluated explicitly and the viscous term implicitly. A simple first-order backward-Euler IMEX discretization is used to evolve equation \ref{eq:mat_vec_second_moment_evolution},
\begin{equation}\label{eq:imex}
    (\boldsymbol{I}-\Delta t \boldsymbol{L})\boldsymbol{f}^{n+1}=\boldsymbol{f}^n-2 \Delta t\Bigl[3\boldsymbol{I} + \boldsymbol{R}\boldsymbol{D}\Bigr]^{-1}\boldsymbol{s}^n.
\end{equation}

\subsection{Linearized solution}
\label{sec:linearized solution}
In spectral space, the linearized EDQNM equation has the analytical solution
\begin{equation}\label{eq:spec_linear}
    \left[\frac{\partial }{\partial t}+2\nu\kappa^2\right]E(\kappa,t)=0\rightarrow E(\kappa,t)=e^{-2\nu\kappa^2t}E(\kappa,0).
\end{equation}
This is analogous to the linear part of the physical space formulation, with the analytical solution given by equation \ref{eq:linear_soln}. To evolve equations \ref{eq:spec_linear} and \ref{eq:linear_soln}, an initial energy spectrum $E(\kappa)$ must be specified. The initial spectrum used is the Batchelor spectrum \citep{batchelor1956large}, defined as
\begin{equation}\label{eq:initial_spectrum}
    E(\kappa)=A\left(\frac{\kappa}{\kappa_p}\right)^m\exp\left[-\beta\left(\frac{\kappa}{\kappa_p}\right)^n\right]
\end{equation}
with $m=4$ for the Batchelor spectrum. Constants are set to $A=32\sqrt{2/\pi}/3$, $\kappa_p=1$, and $\beta=2$. 65 wavenumber points were used which gave an integral-length Reynolds number of $Re_{\ell}=26000$. By transforming the evolved energy spectrum into the longitudinal function using equation \ref{eq:E_to_R}, we can determine how accurately the discrete localized Laplacian performs. Localization of the physical space formulation is dictated by the chosen differentiation matrix. For example, Fourier/Chebyshev collocation is fully nonlocal as the matrix is dense, but the standard 3-point stencil is local as it only requires information from finite neighbors.
\begin{figure}
\centering
\begin{subfigure}{0.45\textwidth}
\centering
\includegraphics[width=\linewidth, trim = 0mm 0mm 0mm 0mm]{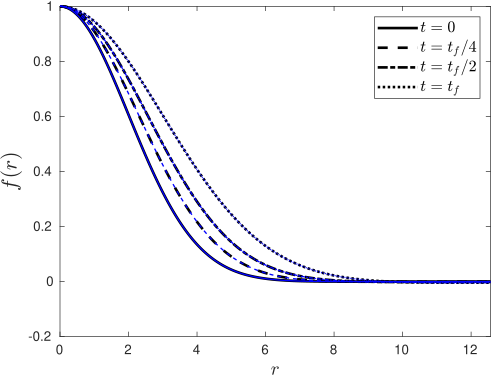}
\caption{Evolution of the longitudinal function $f(r)$. Elapsed time is $t_f=40000$ s.}
\end{subfigure} 
\begin{subfigure}{0.45\textwidth}
\centering
\includegraphics[width=\linewidth, trim = 0mm 0mm 0mm 0mm]{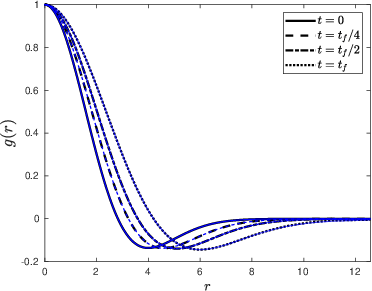}
\caption{Evolution of the longitudinal function $g(r)$. Elapsed time is $t_f=40000$ s.}
\end{subfigure} 
\caption{Change of the longitudinal and lateral functions for various times using the linearized HIT equations with local differentiation matrices. The Batchelor spectrum was used as the initial condition. Blue lines represent the physical space model and black the spectral model.}
\label{fig:linear_localized}
\end{figure}

The longitudinal function was evolved for over $1.3\times 10^4$ eddy turn-over times. Many eddy turn-over times were required to observe a significant change in the shape of $f(r)$. Comparison of the fully spectral and localized physical formulation is given in figure \ref{fig:linear_localized}. The localized physical space evolution is indistinguishable from the nonlocal spectral result, at least for this simple example. Therefore, this shows that the localized physical space formulation is sufficient for this problem. 

\begin{figure}
\centering
\begin{subfigure}{0.45\textwidth}
\centering
\includegraphics[width=\linewidth, trim = 0mm 0mm 0mm 0mm]{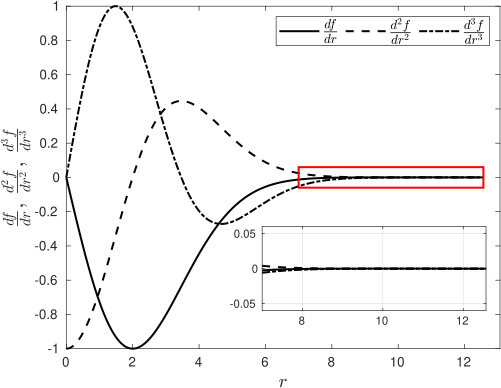}
\caption{Derivatives of $f(r)$  for initial Batchelor spectrum.}
\label{fig:derivatives_batchelor}
\end{subfigure} 
\begin{subfigure}{0.45\textwidth}
\centering
\includegraphics[width=\linewidth, trim = 0mm 0mm 0mm 0mm]{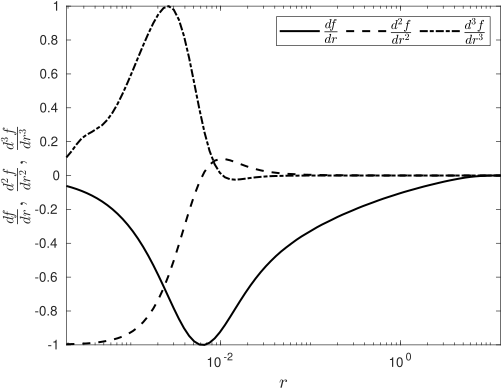}
\caption{Derivatives of $f(r)$ for a realistic spectrum.}
\label{fig:derivatives_realistic}
\end{subfigure} 
\caption{Normalized derivatives of the longitudinal function showing decay at large r-values and sharp features for realistic spectra with a Kolmogorov inertial range.}
\label{fig:derivatives}
\end{figure}

The influence of the finite $r$-distance is also examined. In the full nonlinear problem, closure requires information from $f(r),f'(r),f''(r)$, and $f'''(r)$. It is important to create an $r$-grid that has a sufficiently large maximum $r$-value such that two-point statistics are de-correlated and the Neumann boundary condition is accurate. The drop-off of the derivatives of $f(r)$ for a relatively short maximum $r=4\pi$ is checked in figure \ref{fig:derivatives_batchelor} for a Batchelor initial spectrum, which has longer decorrelation length compared to a realistic spectrum. Figure \ref{fig:derivatives_batchelor} shows that all derivatives are nearly zero at the correlation distance $r=4\pi$. The three normalized derivatives of $f(r)$ are also plotted in figure \ref{fig:derivatives_realistic} for a realistic longitudinal function with a developed inertial range. This $f(r)$ was obtained with EDQNM for $Re_{\ell}=26000$. Note the logarithmic x-axis, showing that most changes occur well below $r=0.1$ This demonstrates the need for geometric $r$-spacing to adequately capture the gradients near $r=0$.

\subsection{Decaying homogeneous isotropic turbulence}
\label{sec:decaying HIT}
Comparison is now made with the spectral EDQNM model for decaying HIT. This verifies that the physical space model evolves to the correct Kolmogorov inertial scaling law and tests invariance to large-scale changes. A large Reynolds number $Re_{\ell}=26000$ is used to ensure the field is turbulent throughout the duration of the simulation.

\begin{figure}
\centering
\begin{subfigure}{0.45\textwidth}
\centering
\includegraphics[width=\linewidth, trim = 0mm 0mm 0mm 0mm]{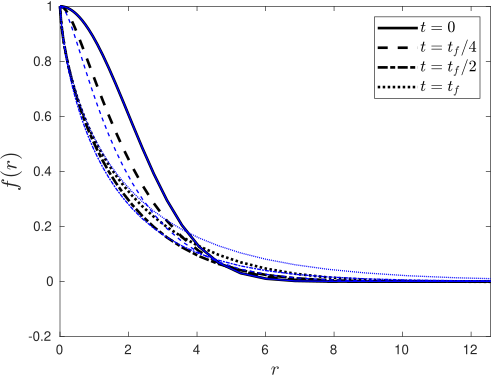}
\caption{Evolution of the longitudinal function $f(r)$. Elapsed time is $t_f=8$ s.}
\end{subfigure} 
\begin{subfigure}{0.45\textwidth}
\centering
\includegraphics[width=\linewidth, trim = 0mm 0mm 0mm 0mm]{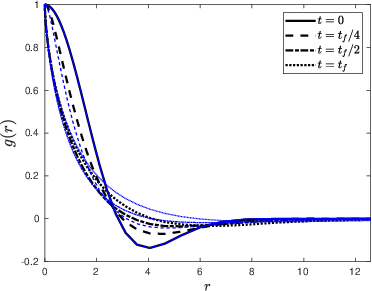}
\caption{Evolution of the longitudinal function $g(r)$. Elapsed time is $t_f=8$ s.}
\end{subfigure} 
\caption{Decaying longitudinal and lateral functions with initial Batchelor spectrum. Blue lines represent the physical space model and black the spectral model. Results are shown for 2.5 eddy turn-over times.}
\label{fig:decaying_funcs}
\end{figure}

\begin{figure}
\centering\includegraphics[width=0.6\linewidth, trim = 0mm 0mm 0mm 0mm]{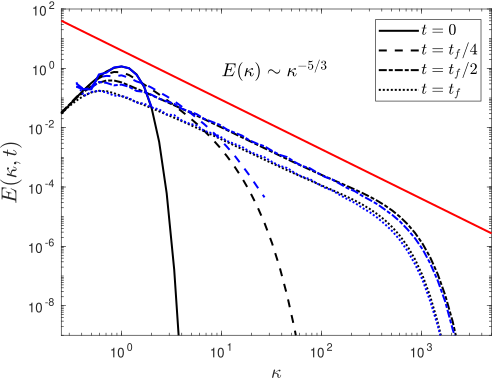}
\caption{Energy spectrum $E(\kappa)$ predictions from the spectral and physical closures for various time intervals starting from the Batchelor spectrum. Blue lines represent the physical space model and black lines the spectral model. Physical space model results are truncated at higher wavenumbers due to non-physical artifacts arising from the transformation.}
\label{fig:decaying_spectrum}
\end{figure}

Figure \ref{fig:decaying_funcs} shows the evolution of the longitudinal and lateral functions $f(r),g(r)$ for 2.5 eddy turn-over times (8 s). Results match well at early and intermediate times, but the physical space model appears to over-predict $f(r)$ and $g(r)$ at larger correlation lengths ($r>2$) when the inertial range is fully present. It is important to note that the main contribution to the dynamics occurs well below this correlation distance (recall figure \ref{fig:derivatives_realistic} shows that the derivatives are mostly non-zero below $r\approx 0.02$), thus the poor prediction at later times at the larger $r$-values do not significantly influence the near $r=0$ quantities and overall evolution of $f(r)$. It was also found that the geometric spacing did not produce much resolution beyond $r>1$ and thus $f(r),g(r)$ evolution displayed more numerical errors in this range. Finally, it is also noted that the Hankel transformation from spectral to physical space (equation \ref{eq:E_to_R}) tends to have more error at larger correlation lengths due to the increased oscillation of the transformation kernel and thus the spectrally-computed $f(r)$ is not a perfect benchmark. Figure \ref{fig:decaying_spectrum} shows the energy spectrum predictions from the spectral and physical models for the same decaying HIT case. Predictions in this space match better, further supporting the observation that the essential regions (such as the inertial regime) are sufficiently captured by the physical space model. Again, due to inaccuracies in the transformation, the energy spectrum computed from the physical space model also has significant numerical error at higher wavenumbers, but this is not shown in figure \ref{fig:decaying_spectrum} because the essential part to compare is the inertial range.

\begin{figure}
\centering
\begin{subfigure}{0.45\textwidth}
\centering
\includegraphics[width=\linewidth, trim = 0mm 0mm 0mm 0mm]{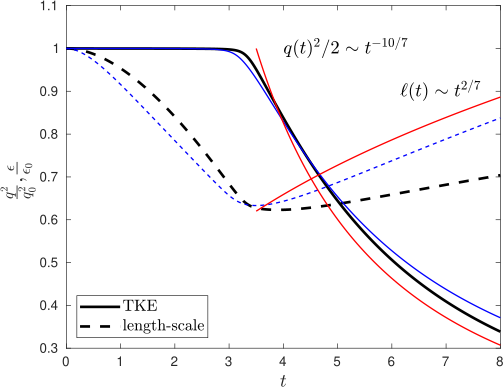}
\caption{Evolution of $q^2(t)/q^2(0)$ and $\ell(t)/\ell(0)$ with initial Batchelor spectrum.}
\end{subfigure} 
\begin{subfigure}{0.45\textwidth}
\centering
\includegraphics[width=\linewidth, trim = 0mm 0mm 0mm 0mm]{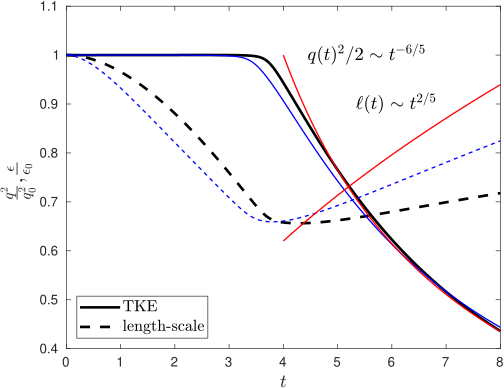}
\caption{Evolution of $q^2(t)/q^2(0)$ and $\ell(t)/\ell(0)$ with initial Saffman spectrum.}
\end{subfigure} 
\caption{Normalized turbulent kinetic energy and integral length scale evolution for decaying HIT with initial Batchelor and Saffamn spectrum. Time scaling laws (red lines) are plotted at $t=3.5$, when the inertial range is sufficiently established. Blue lines represent the physical space model and black the spectral model.}
\label{fig:TKE_decay}
\end{figure}

Long-time turbulent kinetic energy and integral length scale growth are also compared with theory. Different long-time decay laws are verified by changing the initial spectrum. The Saffman and Batchelor spectrums are used, which change the large-scale behavior of the energy spectrum and the implied invariants of the turbulence. Saffman turbulence is initialized using equation \ref{eq:initial_spectrum} with $m=2$, while the Batchelor spectrum uses $m=4$. With Saffman turbulence, the large-scale spectrum behaves as $E(\kappa) \sim \kappa^2 \text{ as } \kappa \to 0$ \citep{saffman1967large}. This corresponds to turbulence with nonzero linear momentum i.e. $\int R_{ij}(\mathbf{r}) \, d^3 r\neq 0$ and represents turbulence generated by flows with large coherent structures or wakes (e.g. grid turbulence). The expected decay laws for the turbulent kinetic energy and integral length scale are $q(t)^2/2 \sim t^{-3/2}, \quad \ell(t) \sim t^{2/5}$ \citep{saffman1967large}. Batchelor turbulence has a different large-scale behavior; $E(\kappa) \sim \kappa^4 \text{ as } \kappa \to 0$. It corresponds to turbulence with zero linear momentum and less memory of the initial momentum. The time decay laws are $q(t)^2/2 \sim t^{-5/2}, \quad \ell(t) \sim t^{2/7}$ \citep{batchelor1956large}. These decay laws are for medium-time behavior, once the universal inertial range is sufficiency established. Figure \ref{fig:TKE_decay} shows the evolution of the normalized turbulent kinetic energy $q^2(t)/q^2(0)$ and integral length scale $\ell(t)/\ell(0)$ for both Batchelor and Saffman turbulence. The traditional spectral EDQNM closure and the physical space closure agree well with respect to the turbulent kinetic energy evolution. Both models also agree on the transition point where the inertial range begins to develop, $t\approx3.5$. There is some mismatch between both models and theory for the length scale growth after the inertial range is established. However, it does appear that the physical space model follows the $\ell(t)\sim t^{2/7}$ more closely than the spectral model.

\subsection{Statistically stationary forced homogeneous isotropic turbulence}
\label{sec:steady forced HIT}
Forced HIT is a most realistic HIT case to verify against because it is reasonable to expect a steady-state solution from the ensemble averaging procedure on which the physical space closure is built. Direct numerical simulation (DNS) data of forced HIT is used as a benchmark to validate the predictive capabilities of both the physical space and traditional EDQNM closures. The DNS data of $1024^3$ with $Re_{\lambda}\approx433$ HIT from the John Hopkins Turbulence Database are used \citep{li2008public,yeung2012dissipation}. The initial spectrum from the database is used to initialize the spectral and physical closure models, with data interpolated to the corresponding correlation/wavenumber grids.

In forced HIT, energy is injected into the large scales to keep the total energy in the shells $|\kappa|\leq2$ constant. The forcing methodology outlined by \citet{donzis2010resolution} is used. The total energy in the shells is
\begin{equation}\label{eq:band_energy}
    E_{\text{band}}(t)=\int^{\kappa_{f2}}_{\kappa_{f1}}E(\kappa,t)d\kappa,\; \gamma=E_{\text{band}}(0)/E_{\text{band}}(t).
\end{equation}
with the scaling ratio $\gamma$ defined from the initial energy in the band. To keep the energy in the shells constant, the energy spectrum is scaled by $\gamma$ at each time-step after computing the dynamics $dE/dt$ for $\kappa\in[\kappa_{f1},\kappa_{f2}]$. This type of forcing preserves spectrum shape inside the band and isotropy \citep{donzis2010resolution}. Using the transformation in equation \ref{eq:E_to_R}, the integral can be rewritten as matrix-vector multiplication $\boldsymbol{f}=\boldsymbol{W}\boldsymbol{e}$ where $\boldsymbol{W}$ is an $m \times n$ matrix which represents the integration kernel and weights, $\boldsymbol{f}$ is a $m\times1$ column vector, again, representing the longitudinal function $f(r)$ scaled by $u'^2$, and $\boldsymbol{e}$ is a $n\times1$ column vector of the energy spectrum evaluated at each wavenumber. The forced longitudinal function can be discretely represented as $\boldsymbol{f}^{(\textrm{forced})}=\boldsymbol{WYW}^{-1}\boldsymbol{f}$ where $\boldsymbol{Y}$ is a diagonal matrix with $\gamma$ on the diagonals for the elements within the forcing band and unity elsewhere. The scaling matrix $\boldsymbol{Y}$ depends on the energy spectrum, which is related to the longitudinal function through equation \ref{eq:E_to_R}, which can be used to rewrite $E_{\textrm{band}}(t)$ into an integral with respect to the longitudinal function,
\begin{equation}\label{eq:E_band_phys}
  E_{\textrm{band}} = \frac{u'^2}{\pi}\int_0^\infty (3f(r)+rf'(r))\Big[ -k_2\cos(k_2 r)+k_1\cos(k_1 r)+\frac{\sin(k_2 r)-\sin(k_1 r)}{r}\Big]dr.
\end{equation}
It was found that this conversion approach was prone to numerical error. The method involves evaluation of highly oscillatory Hankel integrals, which are prone to numerical inaccuracies at large wavenumbers. When transformed back to physical space, these errors manifest as noise in the reconstructed $u'^2f(r)$. An alternative is to directly force the longitudinal function. Several studies have examined forcing strategies for $f(r)$ \citep{lundgren2003linearly, carroll2014effect, rosales2005linear}. However, the objective of this section is not to develop a physically rigorous forcing model, but rather to reproduce the standard low-wavenumber forcing used in the DNS data for consistency and comparison. For this reason, equation \ref{eq:E_band_phys} is used to to force the energy spectrum, which is then converted back to physical space.

\begin{figure}
\centering
\begin{subfigure}{0.45\textwidth}
\centering
\includegraphics[width=\linewidth, trim = 0mm 0mm 0mm 0mm]{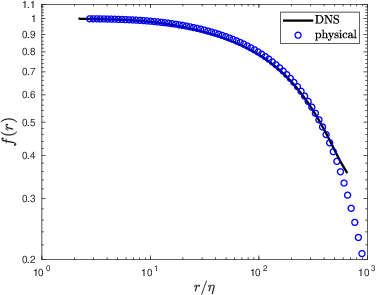}
\caption{Averaged longitudinal function $f(r)$. Elapsed time is $t_f=10$ s or 6 eddy turn-over times.}
\end{subfigure} 
\begin{subfigure}{0.45\textwidth}
\centering
\includegraphics[width=\linewidth, trim = 0mm 0mm 0mm 0mm]{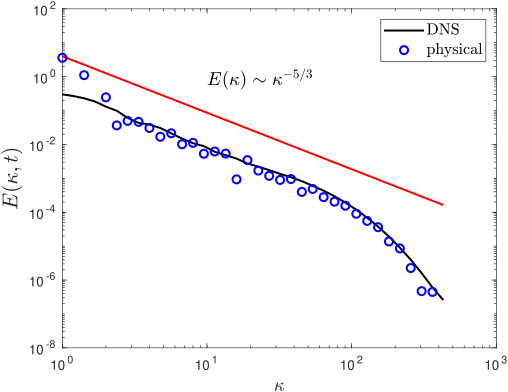}
\caption{Averaged energy spectrum $E(\kappa)$ predictions. . Elapsed time is $t_f=10$ s or 6 eddy turn-over times.}
\end{subfigure} 
\caption{Averaged longitudinal function and energy spectrum with forcing at the large scales. Results are compared to DNS data for the forced HIT at $Re_{\lambda}=433$.}
\label{fig:forced_funcs}
\end{figure}

In figure \ref{fig:forced_funcs} the time-averaged longitudinal and lateral functions are compared to DNS data. The energy spectrum is evolved for $10$ seconds or six eddy turn-over times. Results show that the physical space formulation is able to correctly evolve $f(r)$ and $E(\kappa)$. Results for the longitudinal function are truncated before $f(r)$ reaches zero due to the periodic conditions in DNS, where $f(r)$ computed with the data is not meaningful beyond $L/2=\pi$. This is the reason for the discrepancy of the tail of the DNS $f(r)$, as seen around $r/\eta\approx 5\times10^2$. This demonstrates that the eddy-damping correctly modifies the triple moments and captures the correct Kolmogorov $\kappa^{-5/3}$ scaling in the inertial range, reaching a statistically stationary state.
Higher-order moments and structure functions are analyzed to further probe performance of the nonlinear portion of the formulation. The longitudinal structure functions are also compared for the same datasets. The general expression for the $n^{\text{th}}$-order structure function is
\begin{equation}\label{eq:structure_func}
    S_n(r)\triangleq \langle [(u_i(\boldsymbol{x}+re_i)-u_i(\boldsymbol{x}))e_i]^n\rangle,
\end{equation}
where $e_i$ is a component of a unit vector in an arbitrary direction. K41 theory gives 
\begin{equation}\label{eq:K41_structure_func}
    S_n(r)=C_n(r\epsilon)^{n/3}
\end{equation}
within the inertial subrange. The second-order structure function is connected to the longitudinal function through
\begin{equation}
    S_2(r)=2u'^2(1-f(r)).
\end{equation}
The triple-moment required to close the second-order moment evolution equation, $s(r)= \frac{\partial R_{kii}(\boldsymbol{x,z,y})}{\partial z_k}|_{z=x}$, is studied from various DNS data sets. This third-order moment is computed from the $1024^3$ HIT dataset by averaging across 16 million cells in the domain in the $x$-direction. These two results demonstrate that the energy transfer across scales is correctly captured by the model.

\begin{figure}
\centering\includegraphics[width=0.6\linewidth, trim = 0mm 0mm 0mm 0mm]{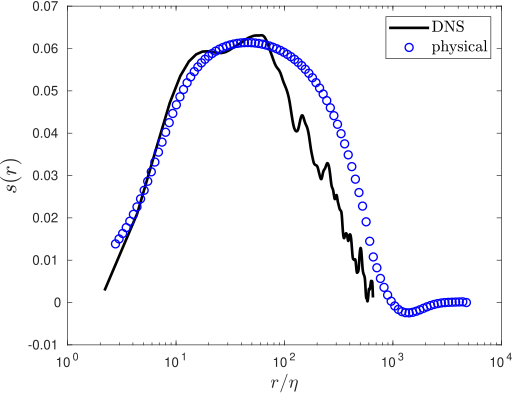}
\caption{Time-averaged two-point triple-moment $\boldsymbol{s}^{(a)}$ of the forced HIT DNS data compared to physical space model predictions. Results are averaged over 10 seconds or approximately 5 eddy turn-over times.}
\label{fig:forced_triple}
\end{figure}
\begin{figure}
\centering\includegraphics[width=0.6\linewidth, trim = 0mm 0mm 0mm 0mm]{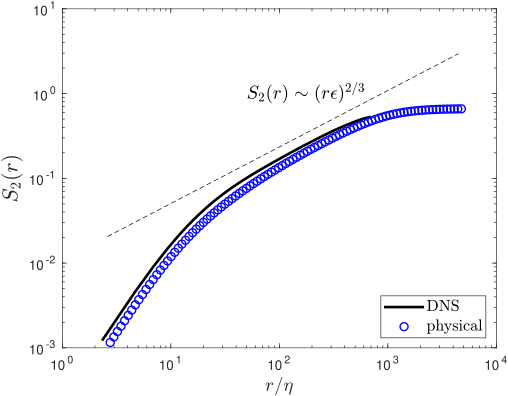}
\caption{Time-averaged second-order structure function $S_2(r)$ of HIT DNS data compared to physical space model. Results are averaged over 10 seconds or approximately 5 eddy turn-over times.}
\label{fig:forced_S2}
\end{figure}
\section{Extension to general problems}
\label{sec:extension to general problems}
The goal of developing the physical space EDQNM formulation was to reduce the barrier towards application to more complex problems. The most immediate issues are that of anisotropy and inhomogeneity. These issues are problematic because the cost of two-point closure becomes significant due to the high-dimensionality of the problem. However, in contrast to traditional spectral EDQNM, the formulation is no longer limited to periodic continuous problems. This section will briefly discuss ideas for extensions to general problems, but the full details and implementation to example problems will be a subject of a separate paper.
\subsection{Anisotropy}
\begin{figure}
    \centering
    \begin{subfigure}{0.45\textwidth}
        \centering
        \includegraphics[width=\linewidth, trim = 0mm 0mm 0mm 0mm]{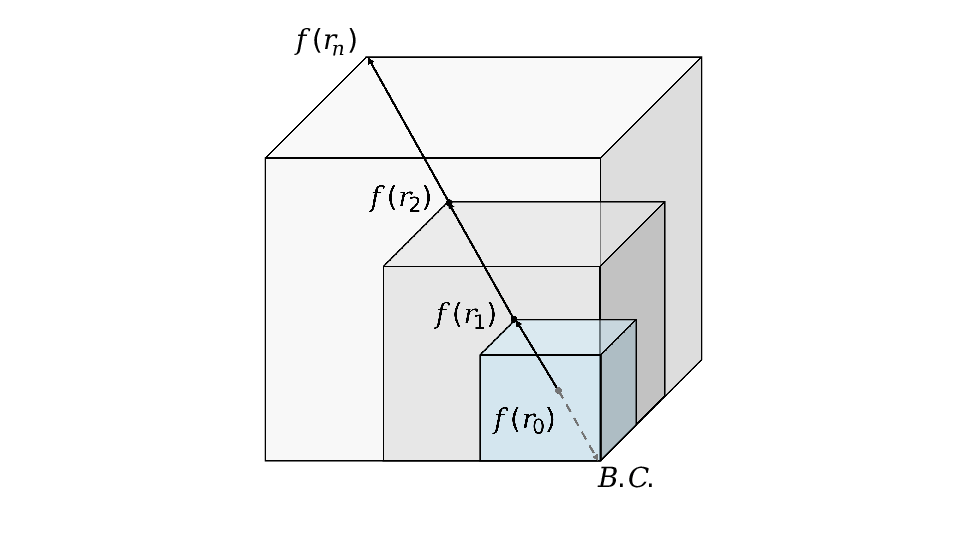}
        \caption{States required for isotropic homogeneous flows.}
        \label{fig:iso_soln}
    \end{subfigure}
    \hfill
    \begin{subfigure}{0.45\textwidth}
    \centering
    \includegraphics[width=\linewidth, trim = 0mm 0mm 0mm 0mm]{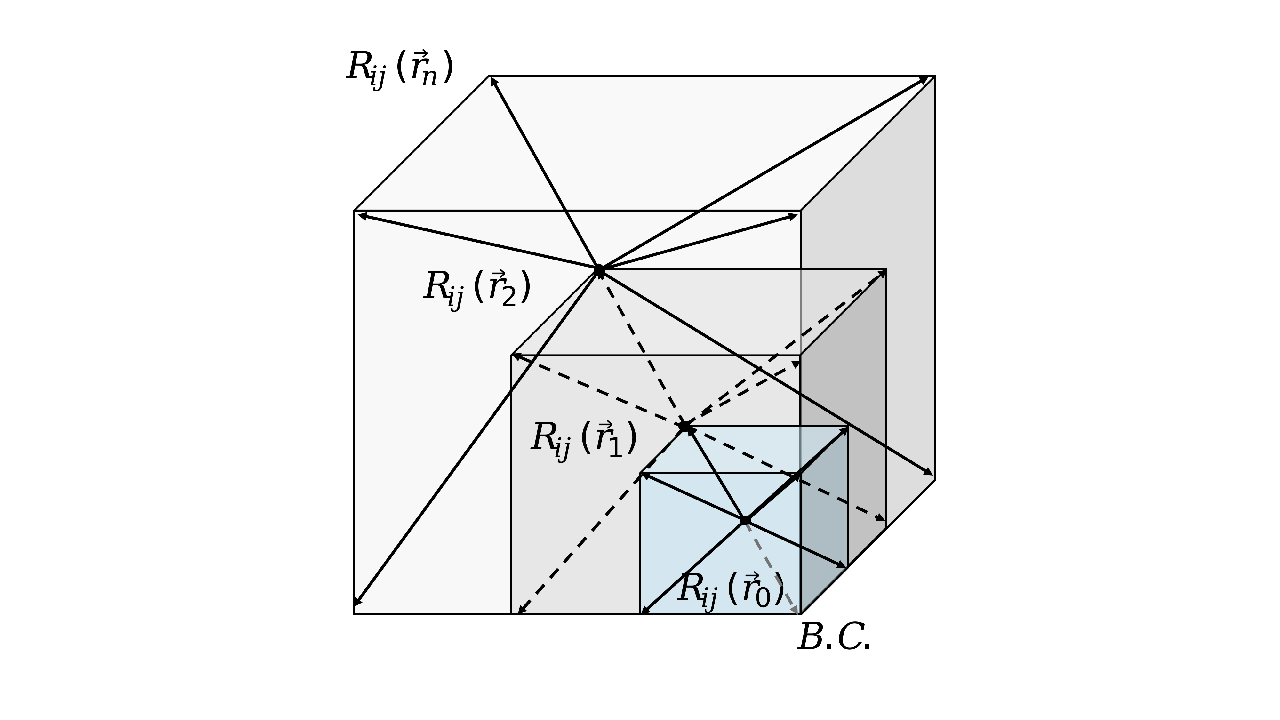}
        \caption{States required for anisotropic homogeneous flows.}
        \label{fig:lateral}
    \end{subfigure}
    \caption{States required for anisotropic homogeneous flows.}
    \label{fig:aniso_solnViso_soln}
\end{figure}

A two-point closure for anisotropic turbulent flow must solve for the direction-dependent Reynolds stress tensor rather than a direction-independent function. This increases to a three-dimensional problem and requires five additional transport equations for each component of the tensor. The solution space for a homogeneous anisotropic flow compared to isotropic flow is compared in figure \ref{fig:aniso_solnViso_soln}.
It is notable that work by \citet{cambon1981spectral}, \citet{cambon2006anisotropic} and others in extending spectral EDQNM to homogeneous anisotropic problems encountered similar issues. Their solution was to decompose the Reynolds stress into isotropic, directional anisotropy, and polarization anisotropy components, reducing the number of evolution equations from six to two. Spherical harmonics were then used to capture the angular dependency. The angular dependency is the main cause for computational cost. To handle the angular dependence, the SO(3) decomposition described by \citet{arad1999correlation} can be used. The SO(3) decomposition is a projection of the correlation tensors onto irreducible representations of the rotation group. It is systematic, orthogonal, and allows one to truncate anisotropy to a few dominant modes.
\begin{equation}\label{eq:SO3_decomp}
    R_{ij}(\mathbf{r}) = \sum_{\ell=0}^\infty \sum_{m=-\ell}^\ell R_{ij}^{(\ell m)}(r) Y_\ell^m(\hat{\mathbf{r}}),
\end{equation}
where $Y_\ell^m$ are spherical harmonics (or basis functions for SO(3)),
$R_{ij}^{(\ell m)}(r)$ are radial coefficients, and $\ell$ indexes the multipole order or “anisotropy level”. For example, $\ell=0$ is the isotropic contribution (spherically symmetric), $\ell=2$ is a quadrupole-like anisotropy (first correction to isotropy), and higher $\ell$ gives finer angular dependence, higher-order anisotropies.

\subsection{Inhomogeneity}
\begin{figure} 
    \centering
    \includegraphics[width=0.7\linewidth, trim = 0mm 0mm 0mm 0mm]{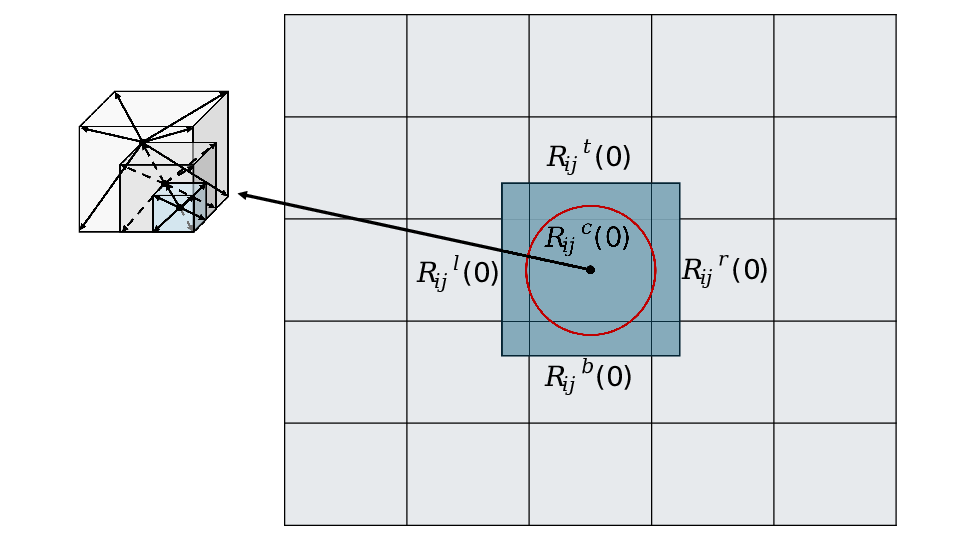}
    \caption{Local 2-point grid for inhomogeneous problem that only requires single-point statistics for the mean flow closure.}
    \label{fig:inhomogeneous_grid}
\end{figure}

In inhomogeneous problems, the two-point correlation is dependent on the absolute spatial location. This is visualized in figure \ref{fig:inhomogeneous_grid}, where the coarse grid is used to store single-point statistics which are used to compute the mean fields. In each coarse grid cell, a local grid is overlaid to compute the two-point statistics. The two-point reach of these local grids may overlap into neighboring coarse grids, as shown by the blue square. The two-point grids must be large enough such that the two-point statistics decay to zero or are approximated in some way such that a boundary condition can be imposed into the differentiation matrices. Computationally, this translates to solving the matrix-vector equations in each coarse grid cell. Depending on how anisotropy is treated, one may also need to solve these matrix-vector equations for an ensemble of directions or evolve transport equations for the SO(3) radial coefficients.
Another point to consider in inhomogeneous problems are the boundary conditions for the two-point correlations. These are non-trivial for points where the correlation distance $r$ reaches a boundary of the overall domain, like a solid wall. In the HIT cases considered in this paper, a Neumann or Dirichlet boundary condition was applied to $f(r)$ as it is expected that $f(r)\to 0$ as $r \to \infty$ due to statistical decorrelation. This is not the case in inhomogeneous problems. For a no-slip wall, for example, we can impose a Dirichlet boundary condition and one-sided stencils for the derivatives. However, if the velocity is non-zero, this simple solution is not applicable. The Reynolds decomposition $u_i=\overline{u}_i+u'_i$ can be used to express the two-point correlation at the boundary as a product of average quantities, i.e. $R_{ij}(\boldsymbol{x}_{BC},\boldsymbol{y})=\overline{u}_i(\boldsymbol{x}_{BC})\overline{u}_j(\boldsymbol{y})+\langle u'_i(\boldsymbol{x}_{BC})u_j(\boldsymbol{y})\rangle$. If the state at the wall is deterministic, then the ensemble of the fluctuating quantity and the state at $\boldsymbol{y}$ is zero. If it is not deterministic, then the boundary two-point correlation must be modeled.

\subsection{Computational complexity}
A quantitative analysis of the additional costs incurred using the physical space framework is now presented. We estimate the computational complexity of the model for HIT and general problems. The cost is computed from the number of floating point operations (flops) required to evolve equation \ref{eq:mat_vec_second_moment_evolution} in one time step. The two input parameters are defined: $N_r$ is number of $r$-grid points and $N_q$ is the number of quadrature points used per integration dimension for the triple integral in the pressure-Poisson equation. In many cases, $N_q\leq N_r$. The fixed parameters of interests are $n_{b}$ average bandwidth size of the differentiation matrices, $n_{s}$ is the flops for the nonlinear integrand evaluation per quadrature point (includes weights, multiplications, adds), $n_{expl}$ is the number of matrix-vector operations required in the explicit terms (approximately 2), and $n_{mv}\approx 2$ is the number of flops per nonzero value for a matrix-vector multiplication. We assume that the implicit solve will use a sparse LU-factorization method with precomputed factorization and a cost of $\mathcal{O}(n_b^2N_r)$. Also, recall that equation \ref{eq:eddy_damping_location} is used to evaluate the nonlinear term $\boldsymbol{s}$. The algorithm by \citet{al2010new} was used to evaluate the sparse matrix exponential-vector product, giving a cost of $\mathcal{O}(n_bN_r)$. The total per-time-step cost for evaluating equation \ref{eq:mat_vec_second_moment_evolution} with an Euler IMEX scheme is
\begin{equation}\label{eq:HIT_cost}
    \text{Cost}_{\text{HIT}} \approx
n_s N_r N_q^{3}+n_{mv} n_{nz} n_{expl} N_r+ 
n_bN_r+n_b^2N_r\sim \mathcal{O}(N_q^3+N_r),
\end{equation}
where the first term is the nonlinear triple/double-integral cost for evaluating the fourth-order moments, the middle term accounts for matrix-vector operations used during evaluation of the explicit parts (viscous term), and the third term is the cost of solving the implicit linear systems and evaluating the matrix exponential for the nonlinear term solution. The first term is dominant if $N_q\approx N_r$. It is also notable that $n_s$ may even be as large as $N_r$ based on the complexity of $M_{a,a}(\boldsymbol{x},\boldsymbol{y})$ and $M_{a,p}(\boldsymbol{x},\boldsymbol{y})$. Cost may be alleviated by reducing the triple integral for $M_{a,p}(\boldsymbol{x},\boldsymbol{y})$ into a double integral, and precomputing parts of $M_{a,p}(\boldsymbol{x},\boldsymbol{y})$ when possible. We note that this computational complexity is unique to incompressible flows.

Computational complexity analysis can be extended to inhomogeneous anisotropic problems. The input parameters are the number of inhomogeneous coarse grid points $N_c$ in each direction and $N_{rc}$ the number of radial coefficients used in the SO(3) group. The fixed constants are $n_t$ the number of moments required to close the mean equations (if only the Reynolds stress is required, there are 6 unique terms so $n_t=6$). Note that in the general problem, the evolution equation will contain derivatives of the two-point moments in arbitrary directions. During each time-step, if SO(3) groups are used, only the derivative of the radial coefficients with respect to separation magnitude needs to be computed; angular derivatives only act on the constant spherical harmonics and can be computed ahead of time and reused. This translates to a constant $n_{sh}$ matrix-vector multiplication at each time-step. The general computational complexity is
\begin{equation}\label{eq:general_cost}
    \text{Cost}_{\text{general}} \approx N_c^3(n_{sh}n_tN_{rc}\text{Cost}_{\text{HIT}})\sim\mathcal{O}(N_c^3N_{rc}(N_q^3+N_r)).
\end{equation}
The extension to anisotropic and inhomogeneous problems each directly scales up the cost. The most expensive is inhomogeneity, as typically $N_c>>N_r,\;N_c>>N_{rc}$. 

Sparse algorithms may be used to reduce computational complexity. One immediate example is reduction of the triple integral using a fast multipole method (FMM). This is accomplished by rewriting the triple integral as a product of pairwise interactions.
\begin{equation}
    \mathcal{F}(r)=\sum^{N_q^2}_{j=1}G(\boldsymbol{r},\boldsymbol{r}_j)q_j,
\end{equation}
where $\mathcal{F}(r)$ is some radial function (Markovian fourth-order moments in this example), $G(\boldsymbol{r},\boldsymbol{r}_j)$ is a kernel that depends only on the relative vector (or can be expanded in multipoles), and $q_j$ are the source strengths. FMM replaces direct summation by hierarchical multipole/local expansions, where instead of a $\mathcal{O}(N_rN_q^3)$ cost, the FMM will have a cost of $\mathcal{O}((N_r+N_q^3)\log p)$ ($p$ is the expansion order). The most significant cost comes from inhomogeneity, where methods for computational cost reduction will come from use of symmetry and adaptive methods. Inhomogeneous grid-size reduction through local refinement can decrease computational cost for problems with sharp features, such as shocks, and use of symmetry can reduce the $N_c^3$ factor to just $N_c$ in some special cases, such as fully-developed boundary layers, etc. Finally, strong emphasis shall be placed on precomputing matrices required for the two-point methods and SO(3) representation. Sparse representation is possible if small localized stencils are used for differentiation matrices and may be tuned for desired accuracy/cost balance. Additional methods for factorized approximations may also be exploited. 

\section{Concluding remarks}
\label{sec:concluding remarks}
A two-point closure model, inspired by the EDQNM model, has been developed entirely in physical space and applied to incompressible homogeneous isotropic turbulence (HIT). The primary objective was to establish a physical space formulation suitable for natural extension to general turbulent flows, enabling the development of predictive closure models. The approach derives evolution equations for arbitrary-order two-point moments directly from the governing dynamical equations, exploiting the linearity of the ensemble averaging operator. This formulation is extendable to both anisotropic and inhomogeneous turbulence.

The resulting two-point evolution equations were expressed in matrix–vector form, revealing how discrete spatial derivatives naturally appear as differentiation matrices. Simplifying assumptions, including quasi-normality and a Markovian approximation, were employed to obtain analytical expressions for the two-point third-order moments in the evolution equation for the velocity correlation. Incorporation of the phenomenological eddy-damping term modeled the effect of fourth-order correlations by introducing a dissipative mechanism acting on the third-order moments.

The physical-space formulation was verified through comparisons with the traditional spectral EDQNM model for decaying HIT and with direct numerical simulation (DNS) data for statistically stationary, forced HIT. Numerical aspects were also examined, particularly the use of a logarithmically-spaced correlation-distance grid to resolve sharp variations near $r=0$, and the implementation of stable algorithms for matrix exponentiation and transformations between physical and spectral spaces.

The limitations of the proposed approach were also assessed regarding its extension to more general flow configurations. The principal challenge lies in the computational expense, as the formulation requires solving a large system of coupled differential equations at each spatial location. The number of equations depends on the representation of anisotropy and the resolution necessary to capture derivatives of the moments with respect to the correlation distance. Some cost-reduction techniques are discussed, such as using FMM for triple-integral evaluations or precomputing spherical harmonic/differentiation matrices used in the two-point evolution and SO(3) representation. Additionally, the coarse-graining operator plays a critical role in determining model behavior; thus, the present formulation is most applicable to statistically stationary problems for which ensemble averaging is well-defined. Further investigation of alternative operators, such as filtering, remains an open direction and may provide a rigorous, physics-based pathway for closure model development.

\backsection[Funding]{Noah Zambrano was supported by the National Science Foundation Graduate Research Fellowship Program. Karthik Duraisamy was supported by the OUSD(RE) Grant \# N00014-21-1-295.}

\backsection[Declaration of interests]{The authors report no conflict of interest.}




\appendix
\section{Pressure term contribution to second-order moment evolution}
\label{sec:negligible pressure}
The use of the pressure Poisson equation to rewrite the pressure in terms of velocity requires the unclosed fourth-order pressure terms to take the form of a nested triple integral\textemdash six integrals total. Fortunately, these terms do not contribute to the evolution of the second-order moment. This is explicitly shown through the pressure terms in the EDQNM closure. The contributions between the advection and pressure terms to the temporal rate of change of the energy spectrum are explicitly found by modifying the projection operator $P_{ijk}(\kappa)$ in the geometrical coefficients appearing in equation \ref{eq:EDQNM_equation}. Advection contributions will have the modified projection operator $P^{(a)}_{ijk}(\boldsymbol{\kappa})=\kappa_j\delta_{ik}+\kappa_k\delta_{ij}$ and the pressure contributions will have $P^{(p)}_{ijk}(\boldsymbol{\kappa})=-\kappa_j\kappa_i\kappa_k/\kappa^2-\kappa_k\kappa_i\kappa_j/\kappa^2$. Upon contraction with the other projection operators, the modified geometrical coefficients $a^{(a)}(\kappa,p,q),\;b^{(a)}(\kappa,p,q)$ are found to be identical to the original geometrical coefficients,
\begin{gather}
    a^{(a)}(\kappa,p,q)=\frac{P^{(a)}_{ijk}(\boldsymbol{\kappa})P_{jl}(\boldsymbol{p})P_{km}(\boldsymbol{q})P_{ilm}(\boldsymbol{\kappa})}{4\kappa^2}=a(\kappa,p,q),\\
    b^{(a)}(\kappa,p,q)=-\frac{P^{(a)}_{ijk}(\boldsymbol{\kappa})P_{jl}(\boldsymbol{q})P_{kil}(\boldsymbol{p})}{2\kappa^2}=b(\kappa,p,q).
\end{gather}
Therefore, the pressure contribution to the nonlinear transfer term must be zero. This property can be verified by generating random $\boldsymbol{\kappa}$ vectors and explicitly solving for the geometrical coefficients, where it is found that the advection-only and total geometrical coefficients always yield the same results.
\section{Quasi-normal fourth-order moment derivations}
\label{sec:markovian term derivations}
The Markovian fourth-order moments (equations \ref{eq:m_a_a}-\ref{eq:markovian_a_p}) appear in both pressure and advection terms in the second-moment evolution equation \ref{eq:second-moment_evolution}. Quasi-normality enables these moments to be recast as products of second-moments. For the pressure terms, Green's function and the pressure Poisson equation must also be used to rewrite the pressure in terms of velocity only.
We now define the distance magnitudes between spatial points as $\boldsymbol{r}\triangleq\boldsymbol{y-x},\boldsymbol{r}'\triangleq\boldsymbol{z-x}$,$\boldsymbol{r}''\triangleq\boldsymbol{w-z}$, and the relative distances as $\boldsymbol{y-z}=\boldsymbol{r-r}'$, $\boldsymbol{y-w}=\boldsymbol{r-r''-r'}$. To help with spherical integration, we define $\boldsymbol{r}$ as the reference vector (any reference direction is valid due to rotational invariance from isotropy),
\begin{equation}
\boldsymbol{r} = r \begin{bmatrix} 1 \\ 0 \\ 0 \end{bmatrix}, \quad
\boldsymbol{r}' = r' \begin{bmatrix} \sin\theta' \cos \phi' \\ \sin\theta' \sin \phi' \\ \cos\theta' \end{bmatrix},\quad \boldsymbol{r}'' = r''\begin{bmatrix} \sin\theta'' \cos \phi'' \\ \sin\theta'' \sin \phi'' \\ \cos\theta'' \end{bmatrix}
\end{equation}
where $\theta$ is the polar angle and $\phi$ is the azimuth. For example, $\boldsymbol{r}'$ aligns with $\boldsymbol{r}$ when $\phi'=0,\theta'=\pi/2$. This gets rid of the cartesian coordinate dependencies and instead gives a solution fully in terms of angles and magnitudes. This makes integration simpler and is given by equation \ref{eq:spherical_integral}. An important note is that terms with mixed distance values also change in magnitude for various directions. This is seen through the magnitude-orientation coupling in the $|\boldsymbol{r}-\boldsymbol{r}'|$ relative distance,
\begin{equation}
    |\boldsymbol{r}-\boldsymbol{r}'|=\sqrt{r^2+r'^2-2rr'\sin{\theta}\cos{\phi}}.
\end{equation}

\subsection{Solving for $M_{a,a}(\boldsymbol{x,y})$}
The goal of this section is to rewrite $M_{a,a}(\boldsymbol{x,y})$, given by equation \ref{eq:markovian_a_a}, in terms of $f(r)$. We start by applying quasi-normality to each of the four terms in equation \ref{eq:markovian_a_a}. Then, after transforming the absolute spatial derivatives to derivatives with respect to $r$, we obtain 
\begin{multline}
    -\frac{\partial ^2R_{jkii}(\boldsymbol{x,z,w,y})}{\partial z_j\partial w_k}\Bigg|_{z,w=x}\approx
 -\frac{\partial R_{jk}(0)}{\partial r_j}\frac{\partial R_{ii}(r)}{\partial r_k}-\frac{\partial R_{ji}(0)}{\partial r_k}\frac{\partial R_{ki}(r)}{\partial r_j}-R_{ji}(r)\frac{\partial^2 R_{ki}(0)}{\partial r_j\partial r_k},
 \end{multline}

\begin{multline}
    -\frac{\partial ^2R_{jiki}(\boldsymbol{x,z,x,y})}{\partial z_j\partial z_k}\Bigg|_{z=x}\approx  -\frac{\partial^2 R_{ji}(0)}{\partial r_j\partial r_k}R_{ki}(r)-R_{jk}(0)\frac{\partial^2 R_{ii}(r)}{\partial r_j\partial r_k}-R_{ji}(r)\frac{\partial^2 R_{ik}(0)}{\partial r_j\partial r_k},
\end{multline}

\begin{multline}
    -\frac{\partial ^2R_{kiji}(\boldsymbol{x,z,w,y})}{\partial z_j\partial w_k}\Bigg|_{z,w=x}\approx-\frac{\partial R_{ki}(0)}{\partial r_j}\frac{\partial R_{ji}(r)}{\partial r_k}-\frac{\partial R_{kj}(0)}{\partial r_k}\frac{\partial R_{ii}(r)}{\partial r_j}-R_{ki}(r)\frac{\partial^2 R_{ij}(0)}{\partial r_j\partial r_k},
\end{multline}

\begin{multline}
    -\frac{\partial ^2R_{kiij}(\boldsymbol{x,w,z,y})}{\partial z_j\partial w_k}\Bigg|_{w=x,z=y}\approx \frac{\partial R_{ki}(0)}{\partial r_k}\frac{\partial R_{ij}(0)}{\partial r_j}+\frac{\partial R_{ki}(r)}{\partial r_j}\frac{\partial R_{ij}(r)}{\partial r_k}+R_{kj}(r)\frac{\partial^2 R_{ii}(r)}{\partial r_j\partial r_k}.
\end{multline}
Many of these cancel out or are zero due to incompressibility. We only require expressions for:
\begin{equation}
    R_{kj}(r)\frac{\partial^2 R_{ii}(r)}{\partial r_j\partial r_k},\;-R_{jk}(0)\frac{\partial^2 R_{ii}(r)}{\partial r_j\partial r_k},\;\frac{\partial R_{ki}(r)}{\partial r_j}\frac{\partial R_{ij}(r)}{\partial r_k}.
\end{equation}
Beginning with the expression for $-R_{kj}(r)\frac{\partial^2 R_{ii}(r)}{\partial r_j\partial r_k}$, we will go through the simplifications step-by-step. The double derivative can be rewritten as
\begin{equation}
    \frac{\partial^2 R_{ii}(r)}{\partial r_k\partial r_l}=u'^2\frac{\partial^2}{\partial r_k\partial r_j}\left[3f(r)+ rf'(r)\right].
\end{equation}
Now, we take derivatives sequentially and obtain,
\begin{equation}
    \frac{\partial^2 R_{ii}(r)}{\partial r_j\partial r_k}=u'^2\left[f'''(r)\frac{r_jr_k}{r}+f''(r)\left(\delta_{jk}+4\frac{r_jr_k}{r^2}\right)+4f'(r)\left(\frac{\delta_{jk}}{r}-\frac{r_jr_k}{r^3}\right)\right].
\end{equation}
Finally, we take the product with $R_{kj}(r)$
\begin{equation}
R_{kj}(r)\frac{\partial^2 R_{ii}(r)}{\partial r_j\partial r_k}=u'^4\left[f(r)\left(rf'''(r)+7f''(r)+\frac{8}{r}f'(r)\right)+f'(r)\left(rf''(r)+4f'(r)\right)\right].
\end{equation}
The next term follows the same process but is multiplied by $R_{jk}(0)$ instead of $R_{kj}(r)$.
\begin{equation}
    -R_{jk}(0)\frac{\partial^2 R_{ii}(r)}{\partial r_j\partial r_k}=-u'^4\left[f'''(r)r+7f''(r)+\frac{8}{r}f'(r)\right]
\end{equation}
Finally, the last term is derived using similar steps,
\begin{multline}
    \frac{\partial R_{ki}(r)}{\partial r_j}=u'^2\Bigg[f'(r)\frac{r_j}{r}\delta_{ki}+f''(r)\left(\frac{r_j\delta_{ki}}{2}-\frac{r_kr_ir_j}{2r^2}\right)\\
    +f'(r)\left(\frac{r_j\delta_{ki}-\delta_{kj}r_i-\delta_{ij}r_k}{2r}+\frac{r_kr_ir_j}{2r^3}\right)\Bigg]\\
    \rightarrow \frac{\partial R_{ki}(r)}{\partial r_j}\frac{\partial R_{ij}(r)}{\partial r_k}=-u'^4\left[r f'(r)f''(r) + \frac{3}{2} f'(r)^2 \right]
\end{multline}
Adding these together gives the final expression for $M_{a,a}(\boldsymbol{x},\boldsymbol{y})$, given in equation \ref{eq:m_a_a}

\subsection{Solving for $M_{a,p}(\boldsymbol{x,y})$}
The expression for $M_{a,p}(\boldsymbol{x,y})$ is in equation \ref{eq:markovian_a_p}. We now apply the pressure Poisson equation and quasi-normality to each of the three terms. The first term becomes
\begin{equation}
    -\frac{1}{\rho} \Bigg\langle\frac{\partial p(\boldsymbol{x})}{\partial x_k} \frac{\partial u_i(\boldsymbol{x})}{\partial x_k}u_i(\boldsymbol{y})\Bigg\rangle=-\int \left\langle \frac{\partial u_l(\boldsymbol{z})}{\partial z_j} \frac{\partial u_j(\boldsymbol{z})}{\partial z_l} \frac{\partial u_i(\boldsymbol{x})}{\partial x_k} u_i(\boldsymbol{y}) \right\rangle \frac{\partial G(\boldsymbol{x} - \boldsymbol{z})}{\partial x_k} \, d^3\boldsymbol{z},\end{equation}
where the derivative of Green's function is
\begin{equation}
    \frac{\partial G(\boldsymbol{x} - \boldsymbol{z})}{\partial x_k}=-\frac{x_k-z_k}{4\pi|\boldsymbol{x-z}|^3}=-\frac{r'_k}{4\pi r'^3}.
    \end{equation}
The fourth-moment is factorized using quasi-normality and converted to derivatives of $R_{ij}(r)$, yielding
\begin{multline}
\left\langle \frac{\partial u_l(\boldsymbol{z})}{\partial z_j} \frac{\partial u_j(\boldsymbol{z})}{\partial z_l} \frac{\partial u_i(\boldsymbol{x})}{\partial x_k} u_i(\boldsymbol{y}) \right\rangle \approx 
\left \langle \frac{\partial u_l(\boldsymbol{z})}{\partial z_j} \frac{\partial u_j(\boldsymbol{z})}{\partial z_l} \right \rangle\left \langle \frac{\partial u_i(\boldsymbol{x})}{\partial x_k} u_i(\boldsymbol{y})\right \rangle+\\
\left \langle \frac{\partial u_l(\boldsymbol{z})}{\partial z_j} \frac{\partial u_i(\boldsymbol{x})}{\partial x_k} \right \rangle\left \langle \frac{\partial u_j(\boldsymbol{z})}{\partial z_l} u_i(\boldsymbol{y})\right \rangle+
\left \langle \frac{\partial u_l(\boldsymbol{z})}{\partial z_j}u_i(\boldsymbol{y})  \right \rangle\left \langle \frac{\partial u_i(\boldsymbol{x})}{\partial x_k} \frac{\partial u_j(\boldsymbol{z})}{\partial z_l}\right \rangle\\
=-\frac{\partial^2 R_{lj}(0)}{\partial r_l\partial r_j}\frac{\partial R_{ii}(r)}{\partial r_k}
+\frac{\partial^2 R_{il}(r')}{\partial r'_j \partial r'_k}\frac{\partial R_{ji}(r-r')}{\partial (r_l-r'_l)}+\frac{\partial R_{li}(r-r')}{\partial (r_j-r'_j)}\frac{ R_{ij}(r')}{\partial r'_k\partial r'_l}
\end{multline}
Now, we do the same for the second term in equation \ref{eq:markovian_a_p}, applying the pressure Poisson equation,
\begin{equation}-\frac{1}{\rho}\Bigg\langle\frac{\partial ^2 p(\boldsymbol{x})}{\partial x_i\partial x_k} u_k(\boldsymbol{x})u_i(\boldsymbol{y})\Bigg\rangle=-\int \left\langle \frac{\partial u_l(\boldsymbol{z})}{\partial z_j} \frac{\partial u_j(\boldsymbol{z})}{\partial z_l} u_k(\boldsymbol{x})u_i(\boldsymbol{y}) \right\rangle \frac{\partial^2 G(\boldsymbol{x} - \boldsymbol{z})}{\partial x_k\partial x_i} \, d^3\boldsymbol{z}.\end{equation}
The second derivative of Green's function is 
\begin{equation}
    \frac{\partial^2 G(\boldsymbol{x} - \boldsymbol{z})}{\partial x_k \partial x_i} = \frac{1}{4\pi} \left( \frac{\delta_{ik}}{|\boldsymbol{x} - \boldsymbol{z}|^3} - 3 \frac{(x_i - z_i)(x_k - z_k)}{|\boldsymbol{x} - \boldsymbol{z}|^5} \right)=\frac{1}{4\pi} \left( \frac{\delta_{ik}}{r'^3} - 3 \frac{(r'_i)(r'_k)}{r'^5} \right).
\end{equation}
Imposing quasi-normality and rewriting in terms of $R_{ij}(r)$ gives the factorization,
\begin{multline}
    \left\langle \frac{\partial u_l(\boldsymbol{z})}{\partial z_j} \frac{\partial u_j(\boldsymbol{z})}{\partial z_l} u_k(\boldsymbol{x})u_i(\boldsymbol{y}) \right\rangle
    \approx \left\langle \frac{\partial u_l(\boldsymbol{z})}{\partial z_j} \frac{\partial u_j(\boldsymbol{z})}{\partial z_l}\right\rangle \left \langle u_k(\boldsymbol{x})u_i(\boldsymbol{y}) \right\rangle\\
    +\left\langle \frac{\partial u_l(\boldsymbol{z})}{\partial z_j}u_k(\boldsymbol{x}) \right\rangle \left \langle \frac{\partial u_j(\boldsymbol{z})}{\partial z_l}u_i(\boldsymbol{y}) \right\rangle
    +\left\langle \frac{\partial u_l(\boldsymbol{z})}{\partial z_j} u_i(\boldsymbol{y})\right\rangle \left \langle u_k(\boldsymbol{x})\frac{\partial u_j(\boldsymbol{z})}{\partial z_l} \right\rangle\\
    =\frac{\partial^2 R_{lj}(0)}{\partial r_j\partial r_l}R_{ki}(r)
    -\frac{\partial R_{kl}(r')}{\partial r'_j}\frac{\partial R_{ji}(r-r')}{\partial (r_l-r'_l)}
    -\frac{\partial R_{li}(r-r')}{\partial (r_j-r'_j)}\frac{\partial R_{kj}(r')}{\partial r'_l}
\end{multline}
Finally, we repeat the same process for the third term of equation \ref{eq:markovian_a_p}. The formal solution for this term is
\begin{equation}
    -\frac{1}{\rho}\Bigg\langle\frac{\partial p(\boldsymbol{y})}{\partial y_i} \frac{\partial u_i(\boldsymbol{x})}{\partial x_k}u_k(\boldsymbol{x})\Bigg\rangle=-\int \left\langle \frac{\partial u_l(\boldsymbol{z})}{\partial z_j} \frac{\partial u_j(\boldsymbol{z})}{\partial z_l} \frac{\partial u_i(\boldsymbol{x})}{\partial x_k}u_k(\boldsymbol{x})\right\rangle \frac{\partial G(\boldsymbol{y} - \boldsymbol{z})}{\partial y_i} \, d^3\boldsymbol{z}.
\end{equation}
The derivative of the Green's function is
\begin{equation}
    \frac{\partial G(\boldsymbol{y} - \boldsymbol{z})}{\partial y_i}=\frac{y_i-z_i}{4\pi|\boldsymbol{y-z}|^3}=\frac{r_i-r'_i}{4\pi (r^2+r'^2 -2\,r\,r'\,\sin\theta\cos\phi)^{3/2}}.
\end{equation}
Imposing quasi-normality and simplifying yields
\begin{multline}
    \left\langle \frac{\partial u_l(\boldsymbol{z})}{\partial z_j} \frac{\partial u_j(\boldsymbol{z})}{\partial z_l} \frac{\partial u_i(\boldsymbol{x})}{\partial x_k}u_k(\boldsymbol{x})\right\rangle\approx \left\langle \frac{\partial u_l(\boldsymbol{z})}{\partial z_j} \frac{\partial u_j(\boldsymbol{z})}{\partial z_l}\right\rangle \left \langle \frac{\partial u_i(\boldsymbol{x})}{\partial x_k}u_k(\boldsymbol{x})\right\rangle\\
    +\left\langle \frac{\partial u_l(\boldsymbol{z})}{\partial z_j}\frac{\partial u_i(\boldsymbol{x})}{\partial x_k} \right\rangle \left \langle \frac{\partial u_j(\boldsymbol{z})}{\partial z_l}u_k(\boldsymbol{x})\right\rangle
    +\left\langle \frac{\partial u_l(\boldsymbol{z})}{\partial z_j} u_k(\boldsymbol{x})\right\rangle \left \langle \frac{\partial u_i(\boldsymbol{x})}{\partial x_k}\frac{\partial u_j(\boldsymbol{z})}{\partial z_l}\right\rangle\\
    =-\frac{\partial^2R_{lj}(0)}{\partial r_j\partial r_l}\frac{\partial R_{ki}(0)}{\partial r_k}
    -\frac{\partial^2R_{il}(r')}{\partial r'_j\partial r'_k}\frac{\partial R_{kj}(r')}{\partial r'_l}
    -\frac{\partial R_{kl}(r')}{\partial r'_j}\frac{\partial^2R_{ij}(r')}{\partial r'_k\partial r'_l}
\end{multline}
Now that all three terms are found, we can multiply them by their respective Green's functions, add them together, convert to expressions of $f(r)$, and finally integrate. The 3D integral is found using a symbolic solver from this point due to the tediousness of the algebra required to convert from expressions of $R_{ij}(r)$ to $f(r)$.
\begin{multline}\label{eq:map}
    M_{a,p}(\boldsymbol{x,y})=-\int_0^\infty\int^{\theta=\pi}_{\theta=0}\int^{\phi=2\pi}_{\phi=0}\Bigg[-f'(r')^{2}\Big(\frac{1}{r'}+\frac{1}{r_{yz}^{3}}+\frac{1}{\pi}+3r' -3ar -1\Big)\\
+f''(r_{yz})f'''(r')\Big(\frac{1-a^{2}}{8} + \frac{r^{2}(1-a^{2})}{8} + \frac{1}{r'} + \frac{1}{r_{yz}^{2}} + \frac{1}{\pi}+ r' - ar\Big)\\
+ f'(r')f''(r')\Big(\frac{1}{r_{yz}^{3}}+\frac{1}{\pi}+ r' - ar -2\Big)\\
+ f'(r_{yz})f''(r')
\frac{(r' - a r)\big(a^{2}r^{2} + 4 a r r' - 3 r^{2} - 2 r'^{2} + 12 r_{yz}^{2}\big)}
{8\pi r'^{2}r_{yz}^{3}}\\
+f'(r')f''(r_{yz})\Bigg[
\frac{3(r' - a r)}{4\pi r'^{3}}
-\frac{2}{r'^{3}r_{yz}^{2}\pi}\Bigg(
\frac{r^{3}}{4}\big(a-2a^{3}\big)
+r^{2}r'\Big(a^{2}-\frac{1}{4}\Big)+\frac{r'^{3}}{4}
-\frac{3a r r'^{2}}{4}
\Bigg)\Bigg]\\
+f''(0)f''(r)\Big(a+ r+ \frac{1}{r'^{2}}+ \frac{1}{\pi}+\frac{3}{4}\Big)\\
+f'(r_{yz})f'''(r')\frac{(r' - a r)\big(2r_{yz}^{2} - r^{2}(1-a^{2})\big)}
{8r'r_{yz}^{3}\pi}\\
+f''(r')f''(r_{yz})\Bigg[
-\frac{,ar^{3} - r^{2}r' - r'^{3} + 3a r r'^{2} - 2 a^{2} r^{2} r'}
{4r'^{2}r_{yz}^{2}\pi}
+\frac{(1-a^{2})r^{2}(r' - a r)}{8r'^{2}r_{yz}^{2},\pi}
\Bigg]\\
+f'(r')f'(r_{yz})\Bigg[
\frac{1}{r'^{3}}+\frac{1}{r_{yz}^{3}}+\frac{1}{\pi}+2+\frac{(r' - a r)\big(2a^{2}r^{2} - 2a r r' - r^{2} + r'^{2} + 5r_{yz}^{2}\big)}
{2\pi r'^{3}r_{yz}^{3}}\Bigg]\\
+f'(r)f''(0)\Bigg(\frac{3a}{r'^{2}\pi} + \frac{3r(1-3a^{2})}{8r'^{3}\pi}\Bigg)
\Bigg]r'^2\sin(\theta)dr'd\theta d\phi
\end{multline}
where $a=\cos(\phi)\sin(\theta)$ and $r_{yz}=\sqrt{r^2+r'^2-2rr'\sin{\theta}\cos{\phi}}$. This is unfortunately the most simplified form. This complexity arises from the use of the longitudinal function rather than $R_{ij}(r)$.

\section{Radial basis functions for differentiation matrices}\label{sec:rbf}
The Radial Basis Function–Finite Difference (RBF-FD) method is a typically mesh-free way of approximating derivatives from scattered data points, but in this work, it is used on structured grids. A smooth function ($f(r)$ in this case) is approximated locally as a linear combination of radial basis functions centered on nearby nodes:
\begin{equation}
    f(r) \approx \sum_{j=1}^{m} w_j , \phi(|r - r_j|) + \sum_{k} a_k p_k(r),
\end{equation}
where $\phi(|r - r_j|)$ is a radial basis function, depending only on distance from node $r_j$, $w_j$ are weights for the RBF part, $p_k(r)$ are polynomial augmentation terms (enforcing polynomial exactness), and $a_k$ are the polynomial coefficients. The polyharmonic spline (PHS) RBF is used here
\begin{equation}
    \phi(s) = |s|^m, \quad \text{with odd } m = 3, 5, 7, \dots
\end{equation}
PHS RBFs are conditionally positive definite and require the polynomial augmentation for stability and consistency. At a target point $r_i$, we select a stencil of nearby points $r_j$. We then enforce that the RBF expansion reproduces the exact derivatives of all basis functions at $r_i$. For the interpolation system, we define
\begin{equation}
A_{pq} = \phi(|r_p - r_q|), \quad P_{pk} = r_p^k,
\end{equation}
for $k = 0, 1, 2, \dots, n_p$ where $n_p$ is the polynomial degree. This is used to form the augmented system,
\begin{equation}
\begin{bmatrix}
A & P \\
P^T & 0
\end{bmatrix}
\begin{bmatrix}
w \\ \lambda
\end{bmatrix}
=
\begin{bmatrix}
\frac{d^n}{dr^n}\phi(|r_i - r_j|) \\
\frac{d^n}{dr^n}p_k(r_i)
\end{bmatrix}.
\end{equation}
The top part enforces the derivative matching of the RBFs, and the bottom enforces polynomial constraints (ensuring the weights exactly differentiate polynomials up to the chosen degree). Solving this linear system for each derivative order $n = 1, 2, 3$ gives differentiation weights $w_j^{(n)}$ for the stencil centered at $r_i$.

At the origin $r=0$, the coordinate system is folded; the physical domain $r \ge 0$ represents both positive and negative values in a 1D Cartesian analogy. Thus, functions are either even $( f(-r) = f(r) )$  (Neumann-type, symmetric), or odd ($f(-r) = -f(r)$) (Dirichlet-type, antisymmetric). To enforce this symmetry, the folded RBFs are defined as
\begin{equation}
  \phi_{\text{even}}(r, r_j) = \phi(|r - r_j|) + \phi(|r + r_j|),\;\phi_{\text{odd}}(r, r_j) = \phi(|r - r_j|) - \phi(|r + r_j|).
\end{equation}
At $r=0$,the even basis produces zero first derivative (Neumann BC) and the odd basis produces zero function value (Dirichlet BC). The folded formulation therefore allows the stencil near $r=0$ to remain well-conditioned and symmetric, avoiding the singularity in $1/r$ terms that would otherwise appear in spherical or cylindrical coordinates.
\bibliographystyle{jfm}
\bibliography{jfm}

\begin{thebibliography}{40}
\expandafter\ifx\csname natexlab\endcsname\relax\def\natexlab#1{#1}\fi
\def\au#1{#1} \def\ed#1{#1} \def\yr#1{#1}\def\at#1{#1}\def\jt#1{\textit{#1}} \def\bt#1{#1}\def\bvol#1{\textbf{#1}} \def\vol#1{#1} \def\pg#1{#1} \def\publ#1{#1}\def\arxiv#1{#1}\def\org#1{#1}\def\st#1{\textit{#1}}

\bibitem[Al-Mohy \& Higham(2010)]{al2010new}
{\sc \au{Al-Mohy, A.H.} \& \au{Higham, N.J.}} \yr{2010}  \at{A new scaling and squaring algorithm for the matrix exponential}.  \jt{J. on Matrix Analysis and App.}  \bvol{31}~(3),  \pg{970--989}.

\bibitem[Andr{\'e} \& Lesieur(1977)]{andre1977influence}
{\sc \au{Andr{\'e}, J.C.} \& \au{Lesieur, M.}} \yr{1977}  \at{Influence of helicity on the evolution of isotropic turbulence at high reynolds number}.  \jt{J. of Fluid Mech.}  \bvol{81}~(1),  \pg{187--207}.

\bibitem[Arad {\em et~al.\/}(1999)Arad, L’vov \& Procaccia]{arad1999correlation}
{\sc \au{Arad, I.}, \au{L’vov, V.S.} \& \au{Procaccia, I.}} \yr{1999}  \at{Correlation functions in isotropic and anisotropic turbulence: The role of the symmetry group}.  \jt{Physical Review E}  \bvol{59}~(6),  \pg{6753}.

\bibitem[Arun {\em et~al.\/}(2021)Arun, Sameen, Srinivasan \& Girimaji]{arun2021scale}
{\sc \au{Arun, S.}, \au{Sameen, A.}, \au{Srinivasan, B.} \& \au{Girimaji, S.S.}} \yr{2021}  \at{Scale-space energy density function transport equation for compressible inhomogeneous turbulent flows}.  \jt{J. of Fluid Mech.}  \bvol{920},  \pg{A31}.

\bibitem[Batchelor \& Proudman(1956)]{batchelor1956large}
{\sc \au{Batchelor, G.K.} \& \au{Proudman, I.}} \yr{1956}  \at{The large-scale structure of homogenous turbulence}.  \jt{Philosophical Transactions of the Royal Society of London. Series A, Mathematical and Physical Sciences}  \bvol{248}~(949),  \pg{369--405}.

\bibitem[Bertoglio {\em et~al.\/}(2001)Bertoglio, Bataille \& Marion]{bertoglio2001two}
{\sc \au{Bertoglio, J.P.}, \au{Bataille, F.} \& \au{Marion, J.D.}} \yr{2001}  \at{Two-point closures for weakly compressible turbulence}.  \jt{Phys. of Fluids}  \bvol{13}~(1),  \pg{290--310}.

\bibitem[Besnard {\em et~al.\/}(1996)Besnard, Harlow, Rauenzahn \& Zemach]{besnard1996spectral}
{\sc \au{Besnard, D.C.}, \au{Harlow, F.H.}, \au{Rauenzahn, R.M.} \& \au{Zemach, C.}} \yr{1996}  \at{Spectral transport model for turbulence}.  \jt{Theo. and Comp. Fluid Dynamics}  \bvol{8},  \pg{1--35}.

\bibitem[Bos \& Bertoglio(2006)]{bos2006single}
{\sc \au{Bos, W.J.T.} \& \au{Bertoglio, J.P.}} \yr{2006}  \at{A single-time two-point closure based on fluid particle displacements}.  \jt{Phys. of Fluids}  \bvol{18}~(3).

\bibitem[Cambon {\em et~al.\/}(1981)Cambon, Jeandel \& Mathieu]{cambon1981spectral}
{\sc \au{Cambon, C.}, \au{Jeandel, D.} \& \au{Mathieu, J.}} \yr{1981}  \at{Spectral modelling of homogeneous non-isotropic turbulence}.  \jt{J. of Fluid Mech.}  \bvol{104},  \pg{247--262}.

\bibitem[Cambon \& Rubinstein(2006)]{cambon2006anisotropic}
{\sc \au{Cambon, C.} \& \au{Rubinstein, R.}} \yr{2006}  \at{Anisotropic developments for homogeneous shear flows}.  \jt{Phys. of Fluids}  \bvol{18}~(8).

\bibitem[Canuto {\em et~al.\/}(1988)Canuto, Hussaini, Quarteroni \& Zang]{canutospectral}
{\sc \au{Canuto, C.}, \au{Hussaini, M.Y.}, \au{Quarteroni, A.} \& \au{Zang, T.A}} \yr{1988} {\em Spectral Methods in Fluid Dynamics\/}.  \publ{Springer-Verlagg, New York}.

\bibitem[Canuto {\em et~al.\/}(2007)Canuto, Quarteroni, Hussaini \& Zang~J.]{canuto2007spectral}
{\sc \au{Canuto, C.}, \au{Quarteroni, A.}, \au{Hussaini, M.Y.} \& \au{Zang~J., Thomas~A.}} \yr{2007} {\em Spectral methods: evolution to complex geometries and applications to fluid dynamics\/}.  \publ{Springer}.

\bibitem[Carroll \& Blanquart(2014)]{carroll2014effect}
{\sc \au{Carroll, P.L.} \& \au{Blanquart, G.}} \yr{2014}  \at{The effect of velocity field forcing techniques on the karman--howarth equation}.  \jt{J. of Turbulence}  \bvol{15}~(7),  \pg{429--448}.

\bibitem[Clark \& Spitz(1995)]{clark1995two}
{\sc \au{Clark, T.T.} \& \au{Spitz, P.B.}} \yr{1995}  \bt{Two-point correlation equations for variable density turbulence}. {\em Tech. Rep.\/}.  \org{Los Alamos National Lab.(LANL), Los Alamos, NM (United States)}.

\bibitem[Djenidi \& Antonia(2021)]{djenidi2021modeling}
{\sc \au{Djenidi, L.} \& \au{Antonia, R.A.}} \yr{2021}  \at{Modeling the third-order velocity structure function in the scaling range at finite reynolds numbers}.  \jt{J. of Mathematical Phys.}  \bvol{62}~(8).

\bibitem[Djenidi \& Antonia(2022)]{djenidi2022karman}
{\sc \au{Djenidi, L.} \& \au{Antonia, R.A.}} \yr{2022}  \at{K{\'a}rm{\'a}n--howarth solutions of homogeneous isotropic turbulence}.  \jt{J. of Fluid Mech.}  \bvol{932},  \pg{A30}.

\bibitem[Domaradzki \& Mellor(1984)]{domaradzki1984simple}
{\sc \au{Domaradzki, J.A.} \& \au{Mellor, G.L.}} \yr{1984}  \at{A simple turbulence closure hypothesis for the triple-velocity correlation functions in homogeneous isotropic turbulence}.  \jt{J. of Fluid Mech.}  \bvol{140},  \pg{45--61}.

\bibitem[Donzis \& Yeung(2010)]{donzis2010resolution}
{\sc \au{Donzis, D.A.} \& \au{Yeung, P.K.}} \yr{2010}  \at{Resolution effects and scaling in numerical simulations of passive scalar mixing in turbulence}.  \jt{Physica D: Nonlinear Phenomena}  \bvol{239}~(14),  \pg{1278--1287}.

\bibitem[Durbin \& Pettersson-Reif(2011)]{durbin2011statistical}
{\sc \au{Durbin, P.A.} \& \au{Pettersson-Reif, B.A.}} \yr{2011} {\em Statistical theory and modeling for turbulent flows\/}.  \publ{Wiley}.

\bibitem[Hamba(2015)]{hamba2015turbulent}
{\sc \au{Hamba, F.}} \yr{2015}  \at{Turbulent energy density and its transport equation in scale space}.  \jt{Phys. of Fluids}  \bvol{27}~(8).

\bibitem[Higham(2005)]{higham2005scaling}
{\sc \au{Higham, N.J.}} \yr{2005}  \at{The scaling and squaring method for the matrix exponential revisited}.  \jt{J. on Matrix Analysis and App.}  \bvol{26}~(4),  \pg{1179--1193}.

\bibitem[Kraichnan(1959)]{kraichnan1959structure}
{\sc \au{Kraichnan, R.H.}} \yr{1959}  \at{The structure of isotropic turbulence at very high reynolds numbers}.  \jt{J. of Fluid Mech.}  \bvol{5}~(4),  \pg{497--543}.

\bibitem[Kraichnan(1971)]{kraichnan1971almost}
{\sc \au{Kraichnan, R.H.}} \yr{1971}  \at{An almost-markovian galilean-invariant turbulence model}.  \jt{J. of Fluid Mech.}  \bvol{47}~(3),  \pg{513--524}.

\bibitem[Leith(1971)]{leith1971atmospheric}
{\sc \au{Leith, C.E.}} \yr{1971}  \at{Atmospheric predictability and two-dimensional turbulence}.  \jt{J. of Atmospheric Sci.}  \bvol{28}~(2),  \pg{145--161}.

\bibitem[Lesieur(2008)]{lesieur2008introduction}
{\sc \au{Lesieur, M.}} \yr{2008} {\em Introduction to turbulence in fluid mechanics\/}.  \publ{Springer}.

\bibitem[Lesieur \& Schertzer(1978)]{lesieur1978amortissement}
{\sc \au{Lesieur, M.} \& \au{Schertzer, D.}} \yr{1978}  \at{Amortissement autosimilaire d'une turbulence {\`a} grand nombre de reynolds}.  \jt{J. de mec.}  \bvol{17}~(4),  \pg{609--646}.

\bibitem[Li {\em et~al.\/}(2008)Li, Perlman, Wan, Yang, Meneveau, Burns, Chen, Szalay \& Eyink]{li2008public}
{\sc \au{Li, Y.}, \au{Perlman, E.}, \au{Wan, M.}, \au{Yang, Y.}, \au{Meneveau, C.}, \au{Burns, R.}, \au{Chen, S.}, \au{Szalay, A.} \& \au{Eyink, G.}} \yr{2008}  \at{A public turbulence database cluster and applications to study lagrangian evolution of velocity increments in turbulence}.  \jt{J. of Turbulence} ~(9),  \pg{N31}.

\bibitem[Lundgren(2003)]{lundgren2003linearly}
{\sc \au{Lundgren, T.~S.}} \yr{2003}  \at{Linearly forced isotropic turbulence}.  \jt{Annual Research Briefs, Center for Turbulence Research, Stanford}  \pg{pp. 461--473}.

\bibitem[Millionschtchikov(1941)]{Millionschtchikov1941}
{\sc \au{Millionschtchikov, M.}} \yr{1941}  \at{On the theory of homogeneous isotropic turbulence}.  \jt{C.R. Akad. Sci}  \pg{p. 615}.

\bibitem[Monin(1959)]{monin1959theory}
{\sc \au{Monin, A.S.}} \yr{1959}  \at{The theory of locally isotropic turbulence}.  \jt{Sov. Phys. Dokl}  \bvol{4}~(271),  \pg{192--215}.

\bibitem[Oberlack \& Peters(1993)]{oberlack1993closure}
{\sc \au{Oberlack, M.} \& \au{Peters, N.}} \yr{1993}  \at{Closure of the two-point correlation equation as a basis for reynolds stress models}.  \jt{Applied Scientific Research}  \bvol{51}~(1),  \pg{533--538}.

\bibitem[Orszag(1970)]{orszag1970analytical}
{\sc \au{Orszag, S.A.}} \yr{1970}  \at{Analytical theories of turbulence}.  \jt{J. of Fluid Mech.}  \bvol{41}~(2),  \pg{363--386}.

\bibitem[Orszag(1973)]{orszag1973statistical}
{\sc \au{Orszag, S.A.}} \yr{1973}  \at{Statistical theory of turbulence}.  \jt{Fluid dynamics}  \pg{pp. 237--374}.

\bibitem[Rosales \& Meneveau(2005)]{rosales2005linear}
{\sc \au{Rosales, C.} \& \au{Meneveau, C.}} \yr{2005}  \at{Linear forcing in numerical simulations of isotropic turbulence: Physical space implementations and convergence properties}.  \jt{Phys. of fluids}  \bvol{17}~(9).

\bibitem[Saffman(1967)]{saffman1967large}
{\sc \au{Saffman, P.G.}} \yr{1967}  \at{The large-scale structure of homogeneous turbulence}.  \jt{J. of Fluid Mech.}  \bvol{27}~(3),  \pg{581--593}.

\bibitem[Sagaut \& Cambon(2018)]{sagaut2008homogeneous}
{\sc \au{Sagaut, P.} \& \au{Cambon, C.}} \yr{2018} {\em Homogeneous turbulence dynamics\/}.  \publ{Springer}.

\bibitem[Talman(1978)]{talman1978numerical}
{\sc \au{Talman, J.D.}} \yr{1978}  \at{Numerical fourier and bessel transforms in logarithmic variables}.  \jt{J. of computational phys.}  \bvol{29}~(1),  \pg{35--48}.

\bibitem[Thiesset {\em et~al.\/}(2013)Thiesset, Antonia, Danaila \& Djenidi]{thiesset2013karman}
{\sc \au{Thiesset, F.}, \au{Antonia, R.A.}, \au{Danaila, L.} \& \au{Djenidi, L.}} \yr{2013}  \at{K{\'a}rm{\'a}n-howarth closure equation on the basis of a universal eddy viscosity}.  \jt{Physical Review E—Statistical, Nonlinear, and Soft Matter Physics}  \bvol{88}~(1),  \pg{011003}.

\bibitem[Von~K\'arm\'an \& Howarth(1938)]{karman1938statistical}
{\sc \au{Von~K\'arm\'an, T.} \& \au{Howarth, L.}} \yr{1938}  \at{On the statistical theory of isotropic turbulence}.  \jt{Proceedings of the Royal Society of London. Series A-Mathematical and Physical Sciences}  \bvol{164}~(917),  \pg{192--215}.

\bibitem[Yeung {\em et~al.\/}(2012)Yeung, Donzis \& Sreenivasan]{yeung2012dissipation}
{\sc \au{Yeung, P.K.}, \au{Donzis, D.A.} \& \au{Sreenivasan, K.R.}} \yr{2012}  \at{Dissipation, enstrophy and pressure statistics in turbulence simulations at high reynolds numbers}.  \jt{J. of Fluid Mech.}  \bvol{700},  \pg{5--15}.

\end{thebibliography}
\end{document}